\documentclass[a4paper,pra,twocolumn,floatfix,longbibliography]{quantumarticle}

\pdfoutput=1
\usepackage{amssymb,amsmath,amstext}
\usepackage{graphicx}
\usepackage{epstopdf}
\usepackage{color}
\usepackage{bm}
\usepackage{appendix}
\usepackage[T1]{fontenc}
\usepackage{bbold}
\usepackage{bbm}
\usepackage{latexsym}
\usepackage[colorlinks=true,citecolor=blue,linkcolor=magenta]{hyperref}

\renewcommand{\vec}[1]{\boldsymbol{#1}}  

\long\def\ca#1\cb{} 

\newcommand{\avg}[1]{\langle #1\rangle }
\newcommand{\ket}[1]{|#1\rangle}               
\newcommand{\bra}[1]{\langle #1|}              
\newcommand{\dya}[1]{\ket{#1}\!\bra{#1}}

\newcommand{\LC}{\mathcal{L}}

\newcommand{\SC}{\mathcal{S}}
\newcommand{\TC}{\mathcal{T}}

\newcommand{\Tr}{{\rm Tr}}

\renewcommand{\leq}{\leqslant}
\newcommand{\mte}[2]{\langle#1|#2|#1\rangle }

\renewcommand{\Re}{\text{Re}}

\renewcommand{\vec}[1]{\boldsymbol{#1}}  

\newcommand*{\id}{\openone}

\newcommand{\noisy}{\text{noisy}}
\newcommand{\exact}{\text{exact}}

\begin{document}

\title{Error mitigation with Clifford quantum-circuit data}

\author{Piotr Czarnik}
\affiliation{Theoretical Division, Los Alamos National Laboratory, Los Alamos, NM 87545, USA.}

\author{Andrew Arrasmith}
\affiliation{Theoretical Division, Los Alamos National Laboratory, Los Alamos, NM 87545, USA.}

\author{Patrick J. Coles}
\affiliation{Theoretical Division, Los Alamos National Laboratory, Los Alamos, NM 87545, USA.}

\author{Lukasz Cincio}
\affiliation{Theoretical Division,  Los Alamos National Laboratory, Los Alamos, NM 87545, USA.}

\begin{abstract}
Achieving near-term quantum advantage will require accurate estimation of quantum observables despite significant hardware noise. For this purpose, we propose a novel, scalable error-mitigation method that applies to gate-based quantum computers. The method generates training data $\{X_i^{\text{noisy}},X_i^{\text{exact}}\}$ via quantum circuits composed largely of Clifford gates, which can be efficiently simulated classically, where $X_i^{\text{noisy}}$ and $X_i^{\text{exact}}$ are noisy and noiseless observables respectively. Fitting a linear ansatz to this data then allows for the prediction of noise-free observables for arbitrary circuits. We analyze the performance of our method versus the number of qubits, circuit depth, and number of non-Clifford gates. We obtain an order-of-magnitude error reduction for a ground-state energy problem on 16 qubits in an IBMQ quantum computer and on a 64-qubit noisy simulator.
\end{abstract}
\maketitle


\section{Introduction}

Currently, one of the great unsolved technological questions is whether near-term quantum computers will be useful for practical applications. These noisy intermediate-scale quantum (NISQ) devices do not have enough qubits or high enough gate fidelities for fault-tolerant quantum error correction~\cite{preskill2018quantum}. Consequently, any observable measured on a NISQ device will have limited accuracy. However, candidate applications such as quantum chemistry require chemical accuracy to beat classical methods~\cite{cao2019quantum,mcardle2020quantum}. Similarly quantum approximate optimization has the potential to beat classical optimization when high accuracy is achieved~\cite{qaoa2014,hadfield2019quantum,Crooks_2018}.

\begin{figure}[t]
    \centering
    \includegraphics[width=0.8\columnwidth]{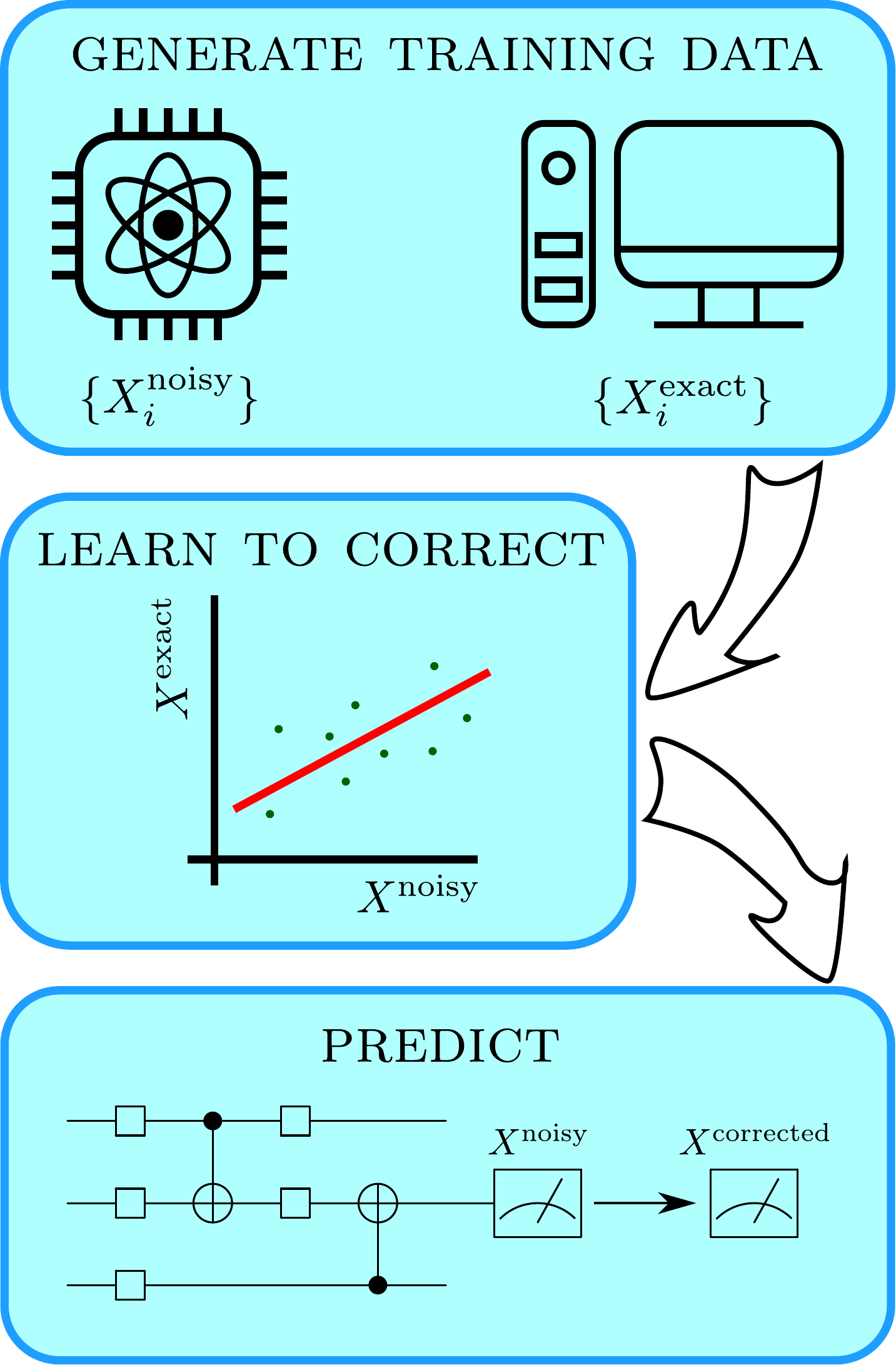}
   \caption{Our proposed error mitigation method. For a set of states that are classically simulable, one generates noisy and corresponding noise-free data on a quantum computer and classical computer, respectively. One learns to correct on this training data by fitting the parameters of an ansatz. Finally, one uses this ansatz with the fitted parameters to predict noise-free observables for arbitrary quantum states.}
    \label{fig:flowchart}
\end{figure}

Hence, it is widely regarded that near-term quantum advantage will only be achieved through error mitigation. Error mitigation (EM) is broadly defined as methods that reduce the impact of noise, rather than directly correct it. EM includes efforts to optimize quantum circuits with compiling and machine learning~\cite{cincio2018learning,murali2019noise,cincio2020machine}. It also includes variational quantum algorithms~\cite{VQE,mcclean2016theory,cerezo2020variationalreview,endo2021hybrid,bharti2021noisy}, some of which can be used to remove the effects of incoherent noise~\cite{Li,QAQC,sharma2019noise,larose2019variational,cerezo2020variational}. A prominent EM approach is to perform classical post-processing of observable expectation values. Perhaps the most widely used example is state-of-the-art zero-noise extrapolation (ZNE), which has shown great promise~\cite{kandala2019error,dumitrescu2018cloud}. ZNE involves collecting data at various levels of noise, achieved by stretching gate times or inserting identities, and using this noisy data to extrapolate an observable's expectation value to the zero-noise limit~\cite{temme2017error,otten2019recovering, endo2018practical, giurgica2020digital, cai2020multi}. It has been successfully employed to correct ground-state energies for problem sizes up to 4-qubits~\cite{kandala2019error,dumitrescu2018cloud,he2020resource}.  Among other approaches are  quasi-probabilistic error mitigation~\cite{temme2017error}  and numerous application specific approaches leveraging symmetries and/or post-selection techniques to mitigate errors~\cite{mcardle2019error,bonet2018low,otten2019noise,cai2021quantum}.

A crucial requirement of any EM method is scalability. While it is relatively easy to develop EM methods for small qubit systems, EM methods that work effectively at the quantum supremacy scale ($> 50$ qubits) are much more challenging to construct. Even methods that are in principle scalable may not actually scale well in practice.

This work aims to address this issue by proposing a novel, EM method that is applicable to all gate-based quantum computers. The basic idea is shown in Fig.~\ref{fig:flowchart}. First we generate training data, of the form $\{X_i^{\noisy},X_i^{\exact}\}$, where $X_i^{\noisy}$ and $X_i^{\exact}$ are the noisy and noiseless versions of an observable's expectation value of interest. The noisy values are obtained directly from the quantum computer, while the noiseless values are simulated on a classical computer. The training data are obtained  from quantum circuits composed largely of Clifford gates (gates that map Pauli operators to Pauli operators), and hence these circuits are efficiently classically simulable. Next we fit the training data with a model, and finally we use the fitted model to predict the noise-free observable.

Our method is conceptually simple and could be refined with sophisticated model fitting methods offered by modern machine learning~\cite{nielsen2015neural}. Nevertheless, even with simple linear-regression-based fitting, our method performs extremely well in practice. We consider the task of estimating the ground-state energy of an Ising spin chain by variationally training the Quantum Alternating Operator Ansatz (QAOA)~\cite{qaoa2014,hadfield2019quantum}. For this task, our method reduces the error by an order-of-magnitude for both a 16-qubit problem solved on IBMQ quantum computer and a 64-qubit problem solved on a noisy simulator, though we expect that this improvement will generically vary depending upon the task considered. We also demonstrate the utility of our method for non-variational algorithms such as quantum phase estimation. Finally we find that our method appears to perform better than ZNE for the larger scale problems we consider.



\section{Our method}
\label{sec:method}
We refer to our method as Clifford Data Regression (CDR). Let $X$ be the observable of interest whose expectation value $X_{\psi}^{\exact}=\mte{\psi}{X}$ one wishes to estimate for a given state $\ket{\psi}$. Let $X_{\psi}^{\noisy}$ be the noisy version of this expectation value obtained from the quantum computer. To remove the noise from this expectation value, the CDR method involves the following steps:

\begin{enumerate}
\item
One chooses a set of states $\SC_{\psi}= \{\ket{\phi_i}\}$ that will be used to construct the training data $\TC_{\psi}$. Each $\ket{\phi_i}$ state must satisfy the property that it is efficient to classically compute the expectation value of $X$ for this state. The CDR method ensures this property by constructing each $\ket{\phi_i}$ state from a quantum circuit composed largely of Clifford gates. We denote the number of non-Clifford gates used to prepare each $\ket{\phi_i}$ as $N$, which (as shown below) plays the role of a refinement parameter.
\item
For each $\ket{\phi_i}\in \SC_{\psi}$, one evaluates $X_{\phi_i}^{\exact} = \mte{\phi_i}{X}$ using a classical computer. One also evaluates the noisy version of this expectation value, $X_{\phi_i}^{\noisy}$, using the quantum computer of interest. These two quantities are incorporated into the training data set $\TC_{\psi} = \{ X_{\phi_i}^{\noisy},X_{\phi_i}^{\exact} \}$.
 
\item 
One constructs an ansatz or model for the noise-free value of the observable in the vicinity of $\ket{\psi}$, 
\begin{equation}
X^{\exact}_{\psi} = f(X^{\noisy}_{\psi}, \vec{a} ),
\label{ansatz}
\end{equation}
where $\vec{a}$ are free parameters. The parameters can be found  either by regression or machine-learning methods. In this article we use least square regression, with a linear ansatz (see Appendix~\ref{Ap:LinearAnsatz} for discussion): 
\begin{equation}
f(X^{\noisy}_{\psi}, \vec{a} )= a_1 X^{\noisy}_{\psi} + a_2,
\label{regression}
\end{equation}
obtaining parameters $a_1$ and $a_2$  by minimizing 
\begin{equation}
C = \sum_{\phi_i \in \SC_{\psi}}  \big(  X^{\exact}_{\phi_i} - ( a_1 X^{\noisy}_{\phi_i}+ a_2)\big)^2.
\label{eqnCostFunction}
\end{equation}
\item 
One uses the ansatz $f(X^{\noisy}_{\psi}, \vec{a} )$ with the fitted parameters to correct $X^{\noisy}_{\psi}$.  
\end{enumerate}

We now discuss several strategies for how to choose $\SC_{\psi}$ in Step 1.  In general, we have found that it is advantageous to tailor the set $\SC_{\psi}$ to the specific state $\ket{\psi}$; in other words, to bias the training data towards the target state of interest. A simple strategy for this purpose is to generate the state-preparation circuits for the $\ket{\phi_i}$ states by replacing a subset of the gates in the circuit that prepares $\ket{\psi}$ with Clifford gates that are close in distance to the original gates. An alternative strategy, which is used in our implementations, is to employ Markov Chain Monte Carlo (MCMC) to generate classically-simulable states $\ket{\phi_i}$ based on the values of their observables. (See Appendix~\ref{Ap:MCMC} for details.) 

In general, one may base the MCMC sampling on the closeness of $X^{\noisy}_{\phi_i}$ to $X^{\noisy}_{\psi}$. However, a potentially more efficient strategy exists for the specific application of variational quantum algorithms~\cite{VQE,mcclean2016theory,yuan2019theory}. When $\ket{\psi}$ is a state that is meant to optimize a variational cost function (e.g., the energy of a given Hamiltonian), then one can instead base the sampling on minimizing this cost function. In this way, one can employ MCMC to obtain classically-simulable states that are close to extremizing the variational cost function.  
For our QAOA implementation (see Section~\ref{sec:QAOA}) we find that this  MCMC scheme produces results that are more robust to implementation details than the simple strategy described above. Furthermore, it outperforms the simple strategy for deep circuits while for shallower ones it  provides results similar  to the best implementations of the simple scheme.      

The shot cost of CDR error mitigation depends on number of circuits in  $\SC_{\psi}$ and the required precision of the $X_{\psi}^{\noisy}$ and $X_{\phi_i}^{\noisy}$ measurements. These quantities need to be determined empirically by analyzing convergence of the results with increasing number of training circuits and shots per 
$X_{\psi}^{\noisy}$ evaluation. In the case of our QAOA implementation  we observe that  $60-80$ training circuits and $8192$ shots per  expectation value estimation are enough to obtain an order of magnitude improvement for all analyzed numbers of qubits $Q=8-64$  and numbers of QAOA rounds $p=1-24$. See details in Section~\ref{sec:QAOA}. 

The classical cost of expectation value computations for near-Clifford circuits  grows exponentially with  $N$ and  polynomially with number of the Clifford gates~\cite{aaronson2004improved}. In the case of state-of-the art classical simulators one can simulate circuits with $N=80$ using a desktop computer ~\cite{pashayan2021fast}. In this work, in order to  provide our method's proof of principle, we use our own implementation  and $N\le30$.  We find  empirically that   number of all near-Clifford circuit simulations  required typically grows polynomially with number of qubits and  circuit depth, see details in Appendix~\ref{Ap:MCMC}.


\section{Numerical Implementations}


\subsection{Implementation for QAOA}
\label{sec:QAOA}

A central application of error mitigation is to correct the energies of Hamiltonian eigenstates prepared on a quantum computer. Here we illustrate this application with the Quantum Alternating Operator Ansatz (QAOA)~\cite{qaoa2014,hadfield2019quantum}, which can serve as an ansatz for Hamiltonian ground states. We consider the transverse Ising chain with open boundary conditions, given by the Hamiltonian
\begin{equation}
H = - g \sum_{j=1}^{Q} \sigma_X^j - \sum_{j=1}^{Q-1} \sigma_Z^j \sigma_Z^{j+1},
\label{H}
\end{equation}
where  $\sigma_X, \sigma_Z$ are Pauli operators, and $Q$ is the length of the chain. We study the case of $g=2$ (belonging to a paramagnetic phase) with open boundary conditions for different numbers of qubits $Q$. To apply the QAOA, we write
$H = H_1 + H_2$ with $H_2 = - g \sum_{j} \sigma_X^j$ and $H_1 =  - \sum_{\langle j,j'\rangle} \sigma_Z^j \sigma_Z^{j'}$. Then the QAOA is 
 \begin{equation}
 \prod_{j=p,p-1\dots,1} e^{i \beta_j H_2} e^{i \gamma_j H_1}  (|+\rangle)^{\otimes Q}, 
 \end{equation}
where $\beta_j, \gamma_j$ are variational parameters, $p$ is the number of ansatz layers, and $\ket{+}=(\ket{0}+\ket{1})/\sqrt{2}$. The exponentials of $H_2$ and $H_1$ can be easily decomposed into quantum circuits (see Appendix~\ref{Ap:QAOA} for a description of the resulting circuits).

 \begin{figure}[t]
\includegraphics[width=\columnwidth]{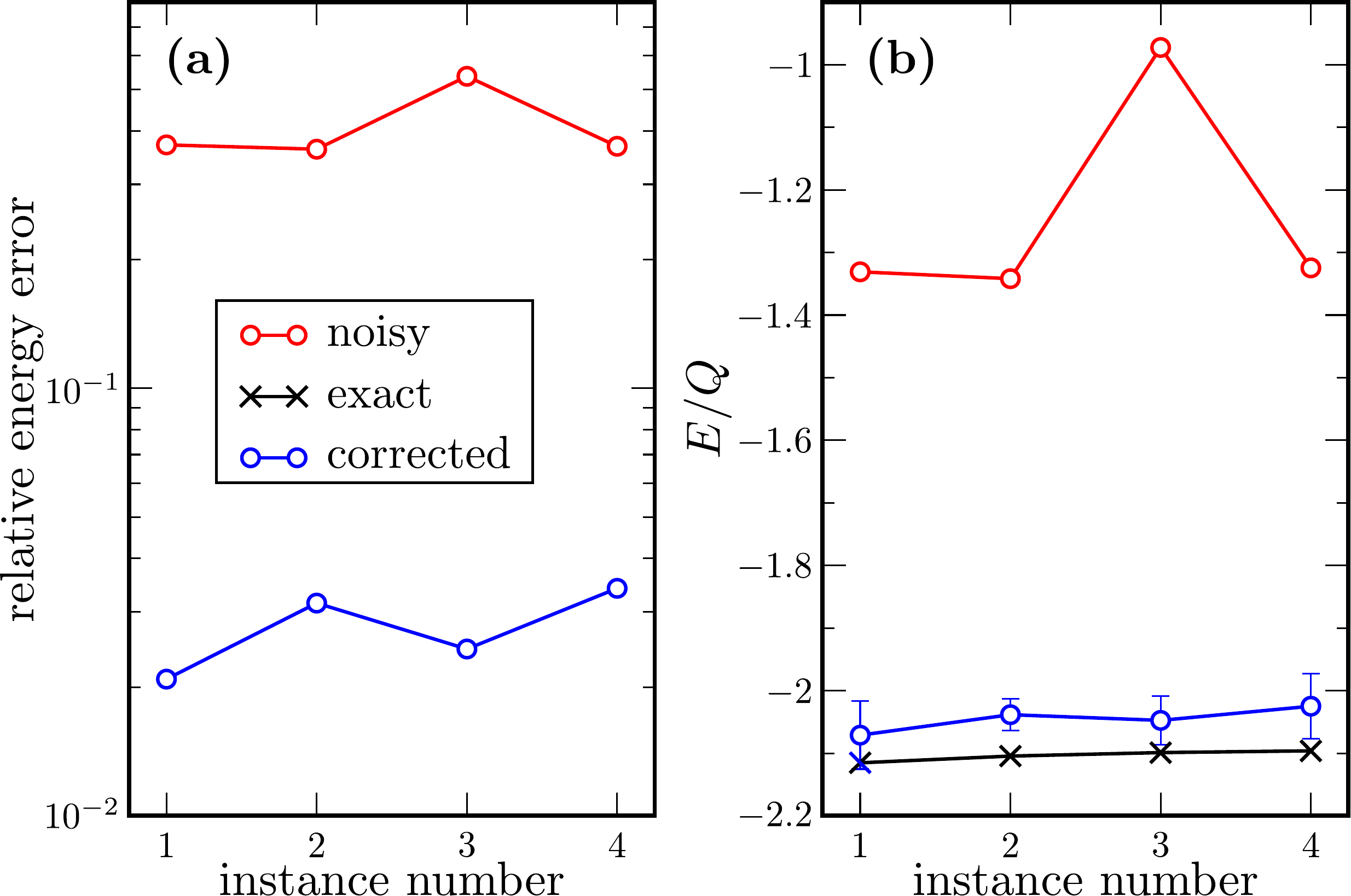}
\caption{ Correcting local minima  of the 16 qubit Ising model energy optimization with  CDR method on IBM's Almaden quantum processor. The mitigated circuits (instances) were prepared by optimizing a QAOA circuit with  $p=2$  rounds. (a) The relative energy error is plotted for the noisy (red) and corrected (blue) results for several optimization instances. (b) The inferred energy per qubit, $E/Q$, is plotted, along with the exact values (black). The error bars are explained in the text. Here the  mean  relative energy error is $0.409$ for the  noisy estimates and  $0.028$ for the corrected ones. The results are obtained with $16384$ shots per circuit. The training sets contain $63$ near-Clifford circuits. Therefore, total number of shots used to mitigate a single instance is $1.05\times10^6$ shots.  }
\label{fig:IBMQresults}
\end{figure}

For this implementation, we perform the optimization of the $\beta_j, \gamma_j$ parameters, and then correct the energies of low-energy local minima of this optimization. This correction involves correcting all $\avg{\sigma_X^j}$ and $\avg{\sigma_Z^{j}\sigma_Z^{j'}}$ terms associated with \eqref{H}. (Note however that the same training set $\SC_\psi$, which is generated based on the total energy, is used to correct each term.) Figure~\ref{fig:IBMQresults} shows that our CDR method reduces the relative energy error (calculated with respect to the exact energy of the minima) by an order-of-magnitude on IBM's Almaden quantum processor. The (recently retired) device had  quantum volume~\cite{cross2019validating} $QV=8$ and connectivity enabling simulation of $Q=16$ Ising chain without use of qubit swap gates.   We remark that the mitigated circuits have $60$ CNOTs and $62$ non-Clifford rotations. We show error bars that reflect our confidence in our ability to correct the training data. Specifically we take the error bars as three standard deviations, where the standard deviation is obtained directly from the cost function in \eqref{eqnCostFunction} by dividing $C$ by the training size and then taking the square root (see Appendix~\ref{Ap:bars} for more details).

\begin{figure}[t!]
\includegraphics[width=\columnwidth]{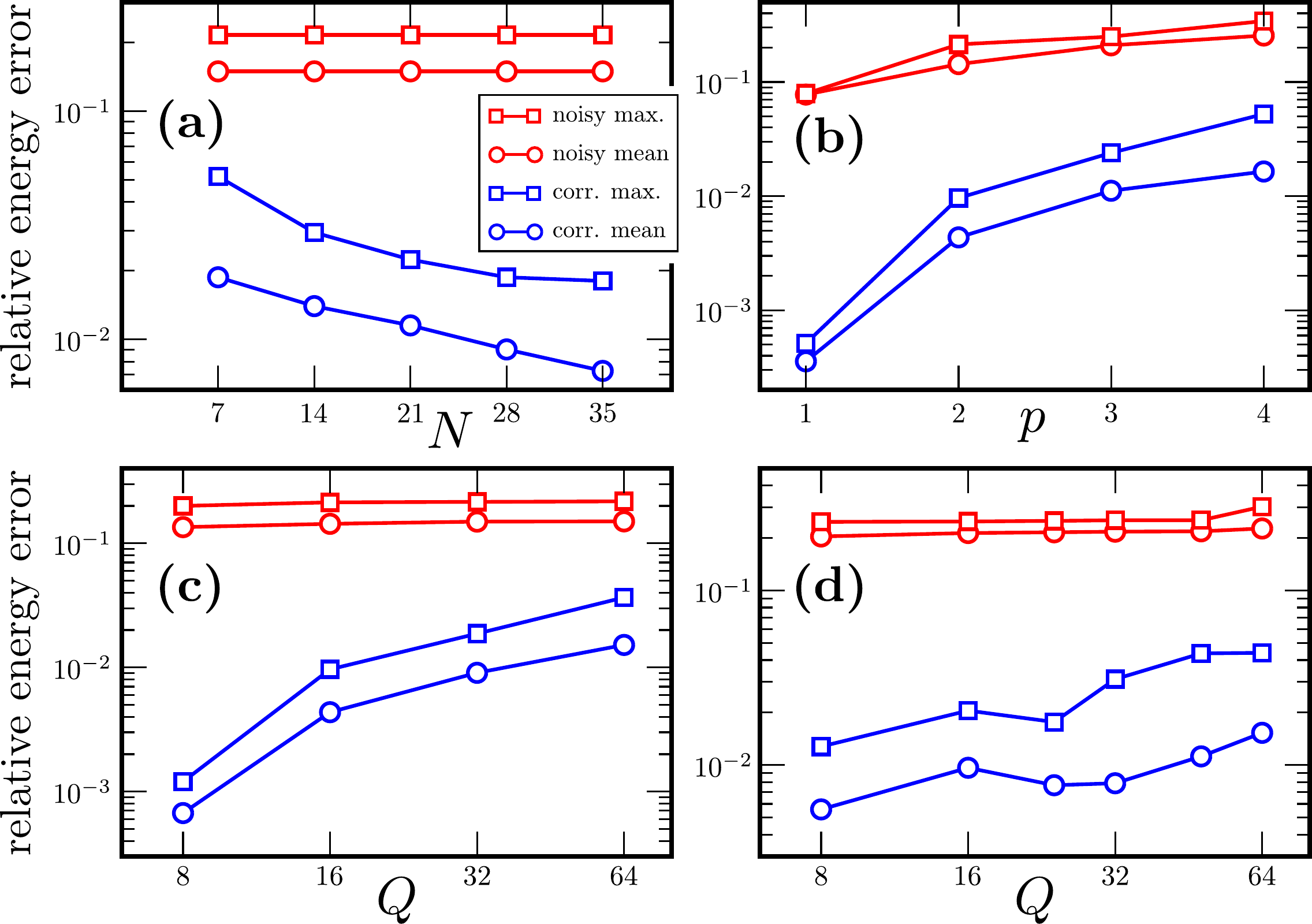}
\caption{Correcting local minima  of the Ising model QAOA energy optimization with our CDR method. The mitigated circuits (instances) were prepared by optimizing a QAOA circuit on a noisy simulator with different  values of initial parameters. The mean (circles) and maximal (squares) relative energy error (calculated with respect to the exact energy of the minima)   is plotted for the noisy (red) and corrected (blue) results. (a) The results for $Q=32,p=2$ plotted versus $N$. (b) The results for $Q=16,N=28$ plotted versus $p$. (c) The results for $p=2,N=28$ plotted versus $Q$. (d) The results for $p=3,N=28$ plotted versus $Q$. Here $Q$ is the number of system qubits, $p$ is the number of QAOA rounds, and $N$ is the number of non-Clifford gates in the training circuits. The noisy results are obtained with $16384$ shots per circuit in the case of (b), (c), (d) and $65536$ shots per circuit in the case of (a). The training sets contain $70$ near-Clifford circuits. The total shot cost of mitigation is $1.16\times10^6$ and $4.65\times10^6$ shots per minimum, respectively. }     
\label{fig:res}
\end{figure}

To study the scaling behavior, we also implement this problem using a classical matrix-product-state~\cite{FannesMPS} simulator and full density matrix  simulators of noisy circuits.  The matrix-product-state simulator enables efficient simulation of  large shallow circuits which are beyond reach of the full density matrix simulator.  
We use a noise model obtained from gate set tomography of IBM's Ourense quantum computer~\cite{cincio2020machine}.
The noise model does not capture spatial and time inhomogeneous aspects of the real device noise. It also does not have cross-talk terms present in the real device. Detailed description of the noise model including its process matrices can be found in~\cite{cincio2020machine}. Furthermore, in our simulations we omit measurement noise as it can be mitigated by specialized techniques~\cite{chow2010detecting,bravyi2021mitigating}.  

Figure~\ref{fig:res} presents the relative energy error, uncorrected (red) and corrected (blue), for different values of $N$, $p$, and $Q$. One can see that our CDR method results in between one to two orders-of-magnitude reduction in the error. Increasing the number of non-Clifford gates $N$ in the training data monotonically reduces the error, as shown in Fig.~\ref{fig:res}(a). This is expected because increasing $N$ allows the training set $\SC_{\psi}$ to become closer to the target state $\ket{\psi}$ of interest. Hence, our results show that $N$ is a refinement parameter, allowing one to obtain better error mitigation with the increased computational difficulty of simulating more non-Clifford gates.  $N$ is limited to $\lesssim 80$ for state-of-the-art classical simulators~\cite{pashayan2021fast}, but our results show that $N\lesssim 30$ already leads to orders-of-magnitude reduction in the error. 

Figures~\ref{fig:res}(b), (c) and (d) show that correcting errors with CDR becomes more challenging with deeper circuits and larger qubit counts, respectively. However we still obtain  order of magnitude error reductions for either $p=4$ layers or 64 qubits. It is worth noting that 64 qubits is considered to be in the regime where quantum advantage might be demonstrated. Furthermore, the $p=4$ ansatz has more two-qubit entangling gates per pair of neighboring qubits than the ansatzes used to demonstrate quantum supremacy~\cite{arute2019quantum}.  We note that the largest circuits considered here ($Q=64,p=3$) contain $378$ CNOTs and $381$ non-Clifford rotations.  

\begin{figure}[t!]
    \centering
    \includegraphics[width=0.99\columnwidth]{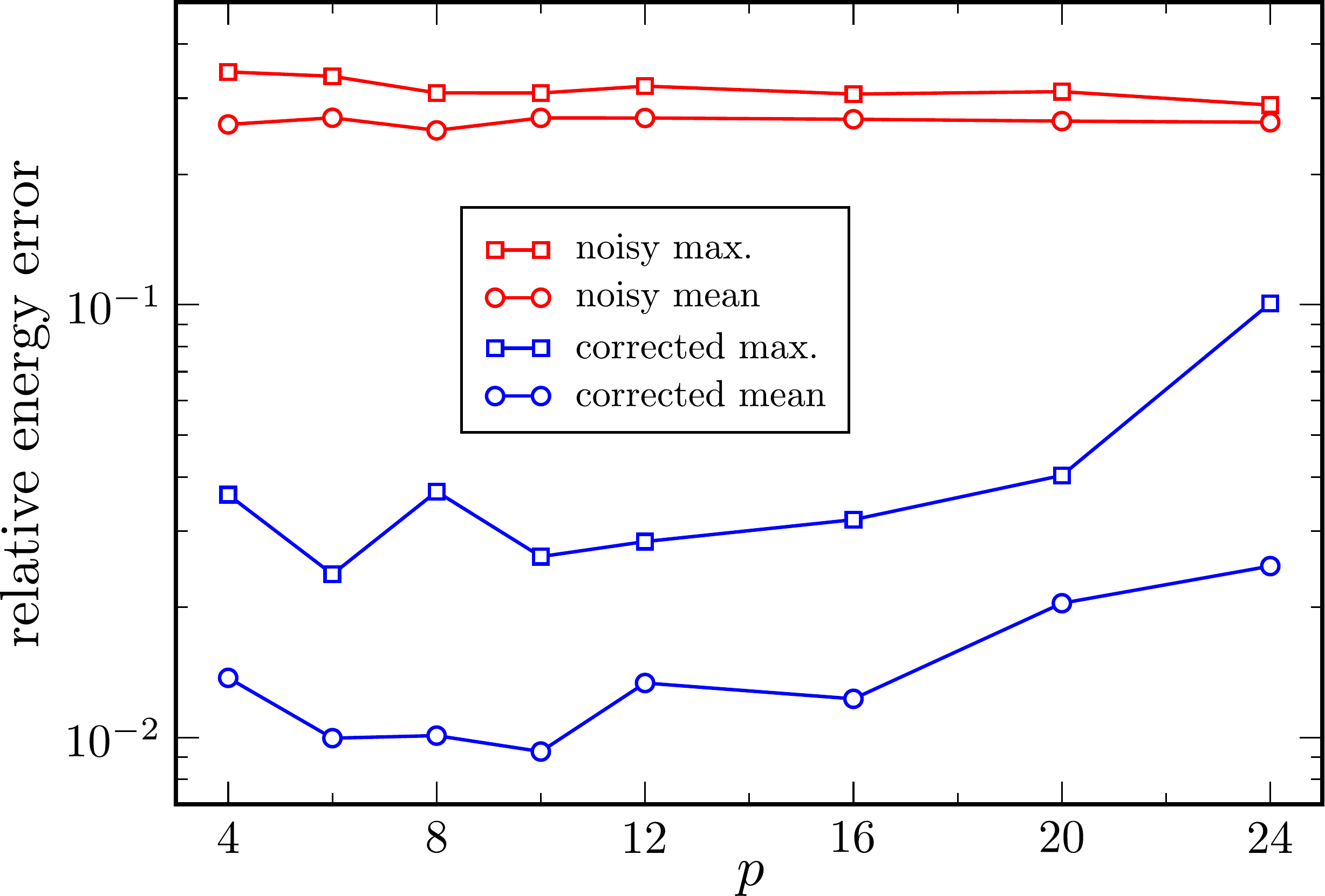}
    \caption{Correcting energies of deep QAOA circuits. Here we show mean and maximal relative energy errors for $30$ local minima of the transverse Ising model energy minimization plotted versus the number of QAOA rounds $p$. The optimization was performed for a system with $Q=8$ qubits using a full density matrix noisy simulator. The red and blue  curves correspond  respectively to   unmitigated  and   CDR mitigated results obtained for  $N=28$ non-Clifford gates in the training circuits.    A noise model used here  has noise rates inversely proportional to $p$ (\ref{noise_rate})  and is  equivalent to the  realistic IBM's Ourense noise model (used in Fig.~\ref{fig:res2}) for $p=4$. The results obtained with $16384$ shots per circuit and $70$ near-Clifford circuits in the training sets. The total shot cost of mitigation is $1.16\times10^6$ shots per circuit of interest.  
    }
    \label{fig:QAOA_deep}
\end{figure}

To obtain substantial improvement  for deeper $p>4$ we consider the case of noise rates  modestly reduced  in comparison to  current devices. To investigate such a scenario we construct  a noise model  by taking gate process matrices $\mathcal{E}$  as   a convex combinations of process matrices of the  IBM's Ourense noise model \cite{cincio2020machine}  $\mathcal{E}_{\textrm{noisy}}$  and process matrices of the noiseless case $\mathcal{E}_{\textrm{exact}}$. 
\begin{equation}
\mathcal{E} = \alpha \mathcal{E}_{\textrm{noisy}} + (1-\alpha)   \mathcal{E}_{\textrm{exact}}, \quad \alpha = \frac{4}{p}. 
\label{noise_rate}
\end{equation}
For this noise model we optimize the QAOA for  $p=4-24$ and  $Q=8$ using full density matrix noisy simulator. Therefore, for $p=4$ we recover the realistic noise model and for the deepest simulated $p=24$  we reduce the noise rates by  a factor of $6$ which is insufficient to enable the fault-tolerant quantum computing~\cite{gottesman2009FTQC}. Again, we mitigate energies of the optimization local minima using  near-Clifford circuits with $N=28$ non-Clifford gates.   We gather results  in Fig.~\ref{fig:QAOA_deep}. The mitigated energies show an order of magnitude improvement even for the deepest $p=24$. Furthermore, we  observe that the improvement remains approximately constant as a function of  number of ansatz layers $p$.  Finally, we note that the deepest circuits simulated here contain $336$ CNOTs and $360$ non-Clifford rotations.  

\begin{figure}[t!]
\includegraphics[width=\columnwidth]{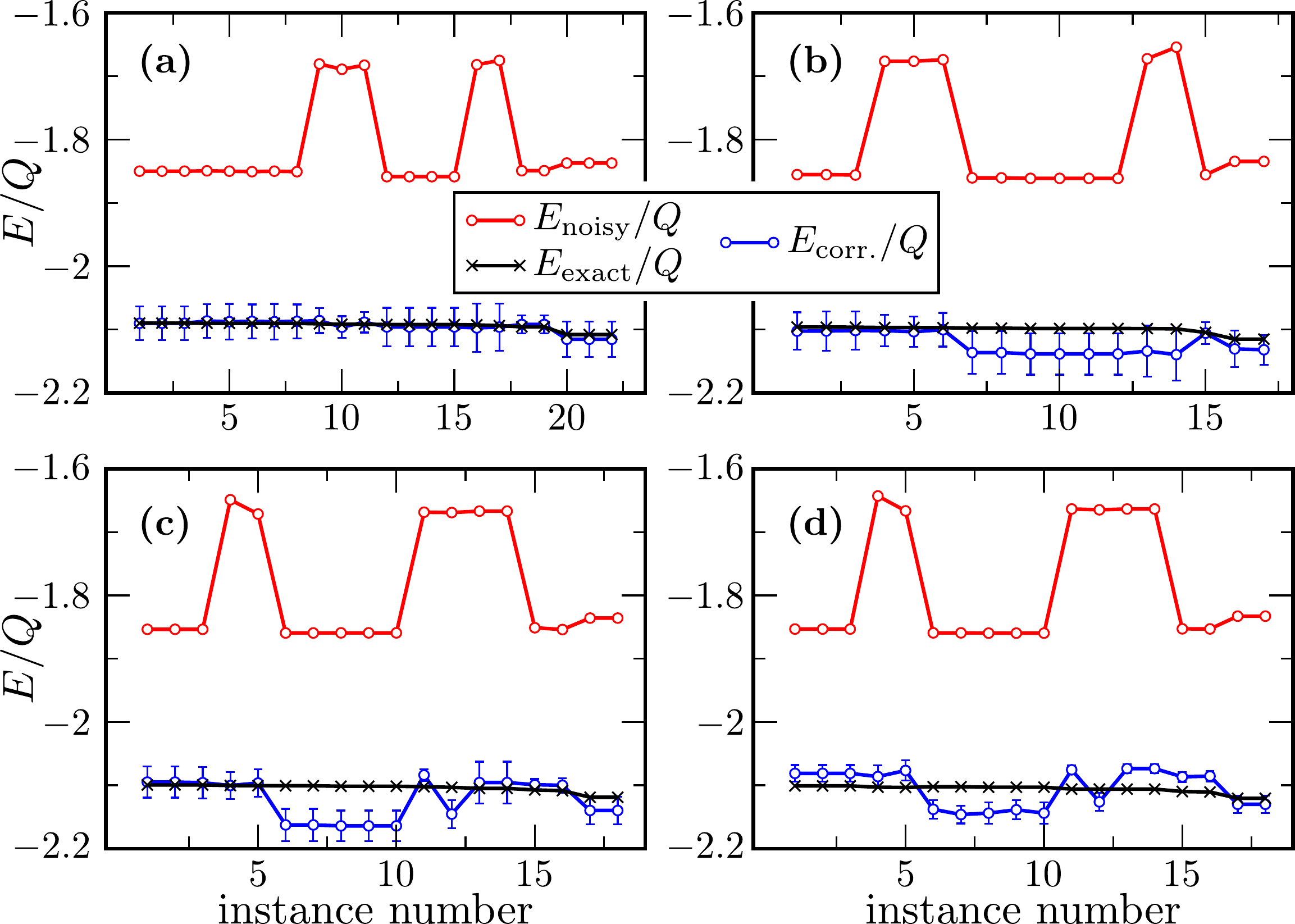}
\caption{Inferred energy per qubit, $E/Q$, for various local minima of  the Ising model  QAOA optimization performed  on a noisy simulator. The noisy, exact, and corrected results correspond to the red, blue, and black curves, respectively. The results were obtained for  QAOA with $p=2$ rounds  for various qubit numbers: (a) $Q=8$, (b) $Q=16$, (c) $Q=32$, and (d) $Q=64$.  CDR was performed  with $N=28$ non-Clifford gates in the training circuits.  The relative errors for these minima were analyzed in Fig.~\ref{fig:res}(c). The error bars are explained in the text.}     \label{fig:res2}
\end{figure}

The scaling results presented in Figs.~\ref{fig:res},~\ref{fig:QAOA_deep} were obtained with $70$ circuits in the training set and $16384$ (Figs.~\ref{fig:res}(b,c,d), ~\ref{fig:QAOA_deep}) or $65536$
(Fig.~\ref{fig:res}(a)) shots per circuit. Therefore, the total shot cost of the CDR mitigation per circuit of interest equals $1.16\times10^6$ and $4.65\times10^6$, respectively. These shot numbers are of the same order as the ones used in our real-device implementations. To perform classical, finite-shot, large $Q$ simulations we use perfect sampling with matrix product states~\cite{ferris2012perfect}.  The training circuits were generated using MCMC sampling with a target energy distribution of the training circuits peaking at an energy close to the ground state energy.  The distribution  was parametrized using the ground state energy, see Appendix~\ref{Ap:MCMC} for details.  Such a distribution favors training circuits with the smallest energy as proposed in Section~\ref{sec:method}. In the case of variational ground state simulations in a quantum advantage regime, the ground state energy is not known and its estimation is one of the goals of the quantum simulation. Therefore, in Appendix~\ref{Ap:MCMC}  we instead use a target distribution which favors low energy training circuits but is not parametrized by the ground state energy. We show numerically that for such a distribution one can obtain very similar results to the ones  shown in  Fig.~\ref{fig:res}. Consequently,
special knowledge of the ground state properties is not necessary to achieve a high quality correction with CDR.

To give further insight into the scaling with $Q$, we show the results for individual optimization instances in Fig.~\ref{fig:res2}. Interestingly, Fig.~\ref{fig:res2} shows that the CDR method is capable of removing noise-induced fluctuations in the energy, i.e., very different noisy energy values are correctly mapped to the same corrected ground-state energy values. For large $Q$, some remnant of these fluctuations still linger in the corrected energies, leading to the worse performance in Fig.~\ref{fig:res}(c), although it suggests that employing a more detailed model in \eqref{ansatz} accounting for additional features in the training data could further improve the accuracy of the corrected  energy.

\subsection{ZNE Implementation}

In addition to applying our CDR method for QAOA, we also examined the performance of the ZNE method under the same conditions as those in Fig.~\ref{fig:IBMQresults}, i.e., for the $Q=16$, $p=2$ case with the same circuits on IBM's Almaden quantum processor.
Furthermore, to mitigate a circuit of interest we used similar total number of shots   ($1.70\times 10^6$) as in the case of our CDR implementation ($1.05\times 10^6$). 
Unfortunately, for this problem instance we were unable to attain a meaningful correction with ZNE.   This difficulty was likely largely due to the number of qubits and depth of the quantum circuit we considered, as ZNE depends on the base circuit being considered not being too noisy to start with. For more details, see Appendix~\ref{Ap:ZNE}.

\subsection{Implementation for phase estimation}
\label{sec:QPE}

\begin{figure}[t!]
\includegraphics[width=\columnwidth,clip=true]{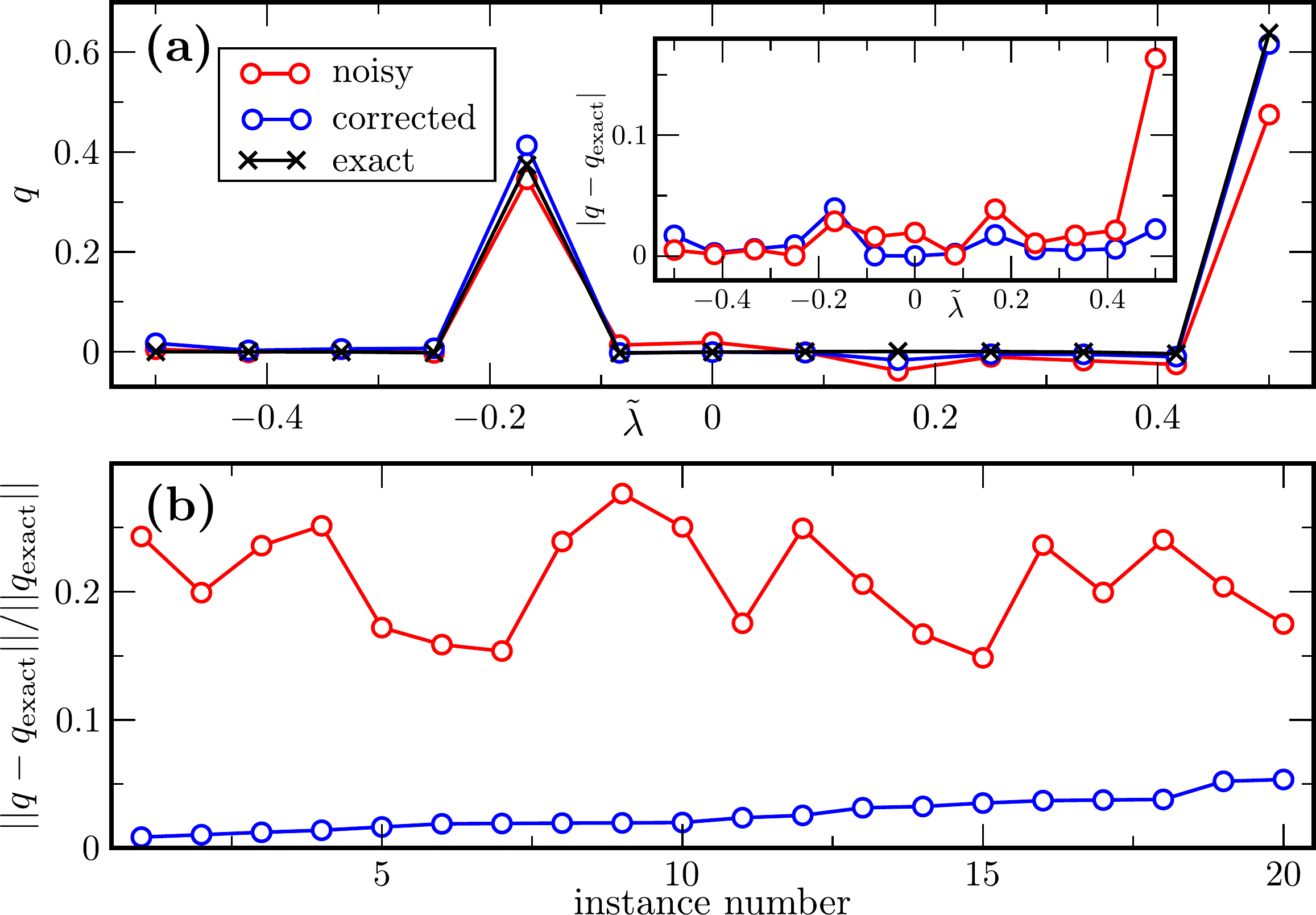}
\caption{Correcting the results of quantum phase estimation on a noisy simulator with CDR. (a) Decomposition  of a random pure state in the binned eigenbasis of $\tilde{H}$ defined in the text, without (red) and with (blue) correction. The probability distribution $q$ is shown as a function of the binned energy eigenvalue $\tilde{\lambda}$. The inset shows its error.  The random state whose decomposition is shown is the one for which CDR performed most poorly. 
 (b) Relative error of this decomposition over 20 instances (different random pure states), ordered by increasing corrected error. For all instances, we employed training data constructed from quantum circuits with $6$ of the total $12$ non-Clifford gates replaced by Clifford gates (see Appendix~\ref{Ap:QPE} for a circuit diagram).}
\label{fig:phestGS}
\end{figure}

Let us now illustrate how our method applies to quantum phase estimation (QPE). We consider a  version of phase estimation  optimized for near-term quantum computers~\cite{somma2019quantum}. For an input state $\ket{\chi}$, this algorithm estimates $\mte{\chi} {e^{-i H t}}=\langle e^{-i H t} \rangle$ on a quantum computer for a series of times $t$ and then classically Fourier transforms the time series to estimate $\ket{\chi}$ decomposition in a binned eigenbasis of $H$. More precisely it  estimates  probability of  $\ket{\chi}$ being in an eigenstate of $H$ with an eigenvalue $\tilde\lambda_j -\epsilon <  \lambda < \tilde\lambda_j+\epsilon$ where $\tilde\lambda_j$ is a center of the bin and $\epsilon$ defines width of the bin with a quantity $q_j$ which definition and convergence properties are discussed in detail in~\cite{somma2019quantum}. The bins are chosen to cover the whole spectrum of $H$.

The expectation value $\langle e^{-i H t} \rangle$ is obtained by measuring $\langle \sigma_X +i \sigma_Y \rangle$ on an ancilla qubit after applying the controlled $e^{-i H t}$ gate to the state $\ket{\chi}$~\cite{somma2019quantum}. 
We performed error mitigation of $\langle e^{-i \tilde{H} t} \rangle$ for a three-qubit Max-Cut Hamiltonian $\tilde{H} = \frac{1}{6} (\sigma^{1}_Z \sigma^{2}_Z +\sigma^{1}_Z \sigma^{3}_Z + \sigma^{2}_Z \sigma^{3}_Z) $, $\ket{\chi}$ being a random pure state, and times $t\in \{1,2,...,136\}$. We calculated $q_j$ for a binned eigenbasis with $\tilde\lambda_j=-0.5,-0.375,\dots,0.5$
using  exact,  noisy and  CDR mitigated $\langle e^{-i H t} \rangle$.  We employed full density matrix simulator using the IBM Ourense noise model~\cite{cincio2020machine} described in more detail in Section~\ref{sec:QAOA}. To reduce circuit depth for our $H$  we apply circuit compilation techniques of ~\cite{cincio2018learning} obtaining a circuit of depth $24$ with $12$ CNOTs and $12$ non-Clifford $R_Z$ rotations, see Appendix~\ref{Ap:QPE} for details. 

 Figure~\ref{fig:phestGS} shows the results. One can see in Fig.~\ref{fig:phestGS}(a) that the decomposition of $\ket{\chi}$ in the energy eigenbasis is significantly improved after correction with the CDR method. Indeed, Fig.~\ref{fig:phestGS}(b) shows that the relative error is reduced by at least a factor of three by CDR.

The QPE algorithm results in circuits with depth growing proportionally to number of $H$ terms which in turn  usually scales polynomially with the number of qubits $Q$.  Consequently,  we expect that  implementation  of the algorithm  for large systems with near-term quantum computers and successful  error mitigation of the resulting circuits will be  challenging despite the algorithm being more resource efficient than its alternatives.  Additionally, circuit compilation techniques used here have cost scaling exponentially with the numbers of qubits~\cite{cincio2018learning} restricting their usability for large systems. Therefore, we do not analyze performance of CDR as a function of $Q$ here.

\section{Conclusions}

With quantum supremacy recently demonstrated~\cite{arute2019quantum}, the next milestone in quantum computing may be to achieve quantum advantage for practical applications such as chemistry or optimization. These applications will require accurate estimation of observables on noisy quantum hardware, and hence large-scale error mitigation will be necessary. In this work, we proposed a  method to significantly reduce the errors (potentially by orders of magnitude) of quantum observables. The method, called Clifford Data Regression (CDR), learns how to correct errors on a training data set. Constructing this training set exploits the classical simulability of quantum circuits composed largely of Clifford gates. This allows one to apply our method to large problem size. For example, we obtained meaningful corrections with CDR for  a 64-qubit ground-state-energy problem, with the total shot number being feasible with current quantum devices. Additionally, we found that the circuit depth our method can mitigate will be notably increased with modest reductions in hardware noise rates, reaching 24 rounds of 8-qubit QAOA.

Further testing our method on real quantum hardware will be important, as will be further studying the scaling of our method's performance with circuit depth and problem size. Furthermore, additional benchmarks will be  required to determine the best training set construction methods for circuits generated by non-variational algorithms. In addition, refining our method with ansatzes that are more sophististicated than our linear ansatz (e.g., using neural networks~\cite{torlai2019precise} or other machine-learning approaches) could be fruitful. Finally, it would be interesting to extend the applicability of our approach to analog quantum simulators. Furthermore, we note that  a related, but distinct, approach using Clifford circuits to learn quasi-probabilistic error mitigation was recently proposed~\cite{strikis2020learning}.

\section{Acknowledgements}

We thank Kipton Barros, Kenneth Rudinger, and Mohan Sarovar for helpful conversations.  We thank IBM for providing access to their quantum computers. The views expressed are those of the authors and do not reflect those of IBM. All authors acknowledge support from LANL's Laboratory Directed Research and Development (LDRD) program. PJC acknowledges support from the LANL ASC Beyond Moore's Law project. This work was also supported by the U.S. Department of Energy (DOE), Office of Science, Office of Advanced Scientific Computing Research, under the Quantum Computing Application Teams~program. This research used quantum computing resources provided by the LANL Institutional Computing Program, which is supported by the U.S. Department of Energy National Nuclear Security Administration under Contract No. 89233218CNA000001.

\appendix

\section{Motivation for linear ansatz}
\label{Ap:LinearAnsatz}

In this work, we assumed essentially the simplest ansatz possible for correcting noisy observables, given by the linear relationship in Eq.~\eqref{regression}. While more complicated ansatzes may improve the performance of CDR, we note here that there does exist theoretical motivation for a linear ansatz.

Consider a depolarizing noise channel that acts globally on all $Q$ qubits. The action of this channel takes the form: 
\begin{equation}
\rho \to (1-p_{\rm err}) \rho + p_{\rm err} \id /d\,,
\end{equation}
where $d$ is the Hilbert-space dimension and $0\leq p\leq 1$. Suppose that this channel acts repeatedly at $m$ different times throughout the quantum circuit of interest. Then, for the circuit that prepares the noise-free state $\ket{\psi}$, the actual final state will be:
\begin{equation}
\rho_{\psi} = (1-p_{\rm err})^m \dya{\psi} + (1-(1-p_{\rm err})^m) \id /d\,.
\end{equation}
Similarly, for a circuit that prepares a state $\ket{\phi_i}$ from the training set $\SC_{\psi}$, the actual final state will be:
\begin{equation}
\rho_{\phi_i} = (1-p_{\rm err})^m \dya{\phi_i} + (1-(1-p_{\rm err})^m) \id /d\,.
\end{equation}
For the observable of interest, $X$, the actual relationship between the noisy and noise-free values is given by
\begin{align}
X_{\psi}^{\noisy} &= \Tr(\rho_{\psi} X) \nonumber\\
&= (1-p_{\rm err})^m X_{\psi}^{\exact} +(1-(1-p_{\rm err})^m)\Tr(X)/d\,,
\end{align}
which is equivalent to
\begin{align}
X_{\psi}^{\exact} = &a_1 X_{\psi}^{\noisy} +a_2 \quad \text{with} \quad a_1=\frac{1}{(1-p_{\rm err})^m}, \nonumber\\ 
& a_2= - \frac{(1-(1-p_{\rm err})^m)\Tr(X)}{d(1-p_{\rm err})^m} \,.
\label{eqn:a1a2}
\end{align}
Hence we see that global depolarizing noise leads precisely to a linear relationship between noisy and noise-free expectation values.

Similarly the training data will take the form of:
\begin{equation}
\{X_{\phi_i}^{\noisy},X_{\phi_i}^{\exact}\}  = \{X_{\phi_i}^{\noisy}, a_1 X_{\phi_i}^{\noisy}+a_2\}\,,
\end{equation}
where $a_1 $ and $a_2 $ are independent of $i$ and are given by the formulas in \eqref{eqn:a1a2}. Fitting a linear regression to the data in the training set will lead to a perfect fit, and the fitted parameters will have the values of $a_1$ and $a_2$ given in \eqref{eqn:a1a2}. Hence, using our linear regression approach in this case will lead to exactly the right correction for $X_{\psi}^{\exact}$.

The above analysis shows that our linear ansatz will perfectly correct for global depolarizing noise. This analysis can be extended to a global version of the quantum erasure channel. More generally, this analysis applies to any channel that maps the input state to a convex combination of the input state and a fixed state, of the form: 
\begin{equation}
    \rho \to (1-p_{\rm err}) \rho + p_{\rm err} \rho_0\,,
    \label{eqn:GeneralNoise}
\end{equation}
where $\rho_0$ is a fixed state.

In addition, we remark 
that the linear ansatz will perfectly correct for certain types of measurement noise. This is because measurement noise typically can be thought of as noise acting on the local state just prior to a noise-free measurement. If this noise corresponds to depolarizing noise (e.g., for symmetric measurement noise) or more generally to the noise in Eq.~\eqref{eqn:GeneralNoise} (which allows for asymmetric measurement noise), then one can apply the analysis above to show that the linear ansatz will perfectly correct the noise.

While we do not expect the linear ansatz to correct for all types of noise, one can nevertheless replace the linear ansatz with a more complicated one, and further work is needed to explore the benefits of such (more complicated) ansatzes.

\section{Generating the set $\SC_{\psi}$}
\label{Ap:MCMC}

\begin{figure*}[t]
    \centering
    \includegraphics[width=0.75\textwidth]{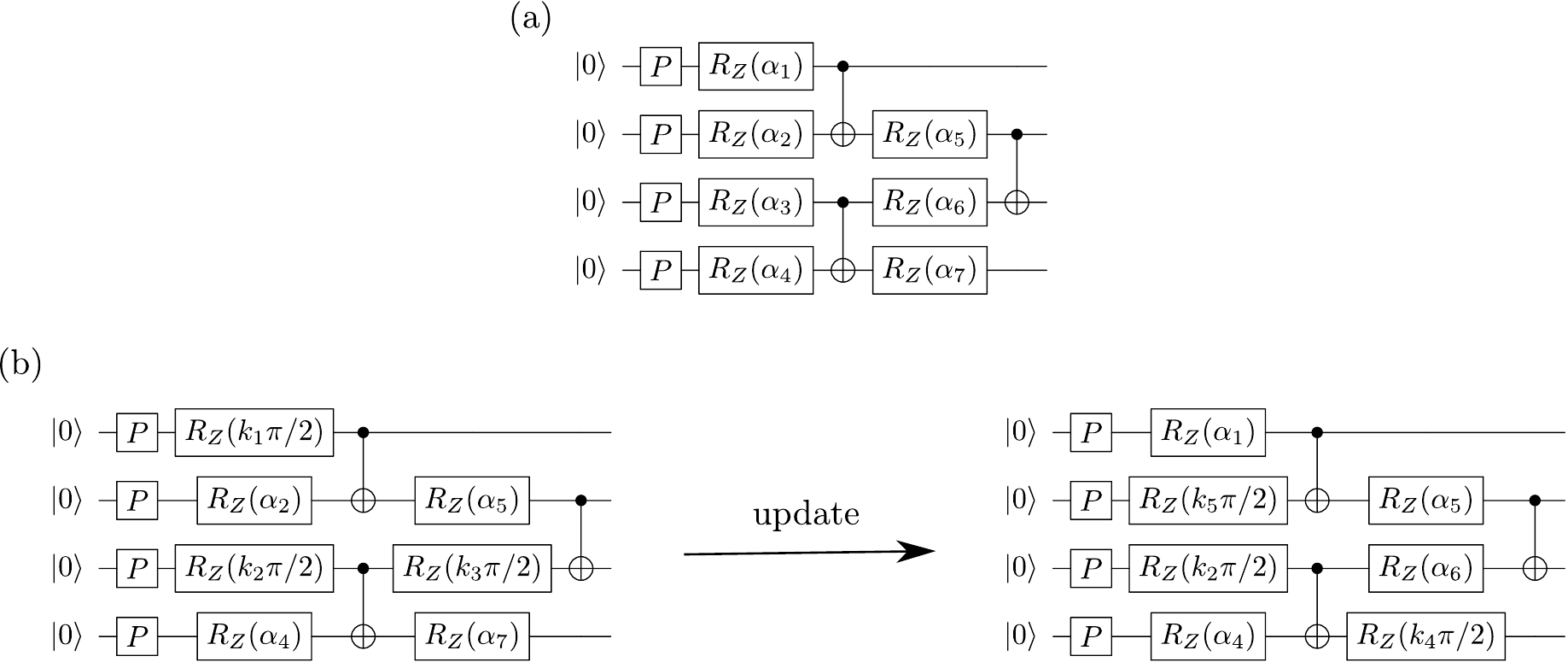}
    \caption{ 
     An  MCMC update used here visualized  for a simple circuit. In (a) we show an exemplary circuit of interest with $7$ non-Clifford gates $R_Z(\alpha_1),\dots, R_Z(\alpha_7)$. In (b) a near-Clifford circuit on the left is updated by choosing $n_p=2$   non-Clifford rotations ($R_Z(\alpha_2), R_Z(\alpha_7)$) and $n_p$ Clifford rotations ($R_Z(k_1 \pi/2), R_Z(k_3 \pi/2)$). The Clifford rotations are replaced by corresponding non-Clifford rotations ($R_Z(\alpha_1), R_Z(\alpha_6)$) of the circuit of interest according to likelihood distribution (\ref{projloc}). The non-Clifford rotations are replaced by original Clifford rotations ($R_Z(k_4 \pi/2), R_Z(k_5 \pi/2)$). The update preserves the number of non-Clifford gates $N$.   As a result of the update we obtain a circuit at the right  which is later accepted or rejected according to the Metropolis-Hastings rule. Here $P=R_X(\pi/2)$ and to obtain Clifford rotations we have $k_1, k_2,\dots,k_7=0,1,2,3$.  
    }
    \label{fig:update_MCMC}
\end{figure*}

We use a Markov Chain Monte Carlo (MCMC)~\cite{van2018simple} sampling technique in order to generate a set of classically simulable training states, $\SC_{\psi}$, for use in CDR. Starting from some initial simulable circuit, we build $\SC_{\psi}$ by making small update steps to our circuit. At each update step, we choose to either accept or reject the change. On accepting a new circuit it is added to $\SC_\psi$, and we look for updates starting from this new circuit. In the case when acceptance of the circuit is based on its exact expectation values (as in the case of our QAOA implementation) one can further improve shot efficiency of the method. As for large circuits it takes many small changes to form a circuit with significant fraction of gates updated one can form  $\SC_{\psi}$ post-selecting the accepted circuits to choose only ones sufficiently different. We formalize this  notion below.   

In our applications, the initial point is chosen by finding a near-Clifford circuit with $N$ non-Clifford gates that is close to the circuit we wish to correct. To generate update steps, we randomly pick $n_p$ pairs of the circuit's $\sigma_Z$ rotations. Here, each pair consists of one rotation that has been replaced by a Clifford gate (specifically, $S^n$ for some integer $n$ where $S = R_Z(\pi/2) = e^{-i\sigma_Z\pi/4}$ is the $\pi/2$ rotation gate) and one that has not. We chose $n_p=5$ in our implementations.
For each pair we then replace the non-Clifford rotation by a power of $S$ and the Clifford gate by the original rotation in that part of the circuit.
When replacing a rotation $R_Z(\alpha)$ by $S^n$, the power $n$ is randomly sampled with weight $w(n)$, which is given by
\begin{equation}
w(n) = e^{-d^2/\sigma^2},\, d = ||R_Z(\alpha) -S^n||,
\label{projloc}
\end{equation} 
with $\sigma=1/2$. Note that such an update preserves the number of non-Clifford gates $N$. We visualize the update for a simple circuit in Fig.~\ref{fig:update_MCMC}.

For our implementations, the new state $\ket{\phi}$ proposed by this update step is then either accepted or rejected according to a Metropolis-Hastings rule with a likelihood function $L$. In the limit of many MCMC steps, such sampling produces a distribution of near-Clifford training circuits with probability proportional to $L$. The likelihood was defined differently for our QAOA and phase estimation examples. In the case of QAOA our goal was to produce probability distribution peaking as close to the ground state energy as possible. 
To that purpose  we used
\begin{equation}\label{eq:QAOA_likelihood}
L(X^{\exact}_{\phi}) \propto e^{-(X^{\exact}-X_0)^2/X_\sigma^2}.
\end{equation}
Here, $X$ is the energy per qubit, and $X_0=-2.1$ being roughly the ground state energy. We used $X_\sigma=0.05$, though we found that instead using $X_\sigma=0.1$ or $0.2$ produced similar results. Larger values ($X_\sigma\sim 1$) degraded the quality of the results. 

\begin{figure}[h!]
    \centering
    \includegraphics[width=0.9\columnwidth]{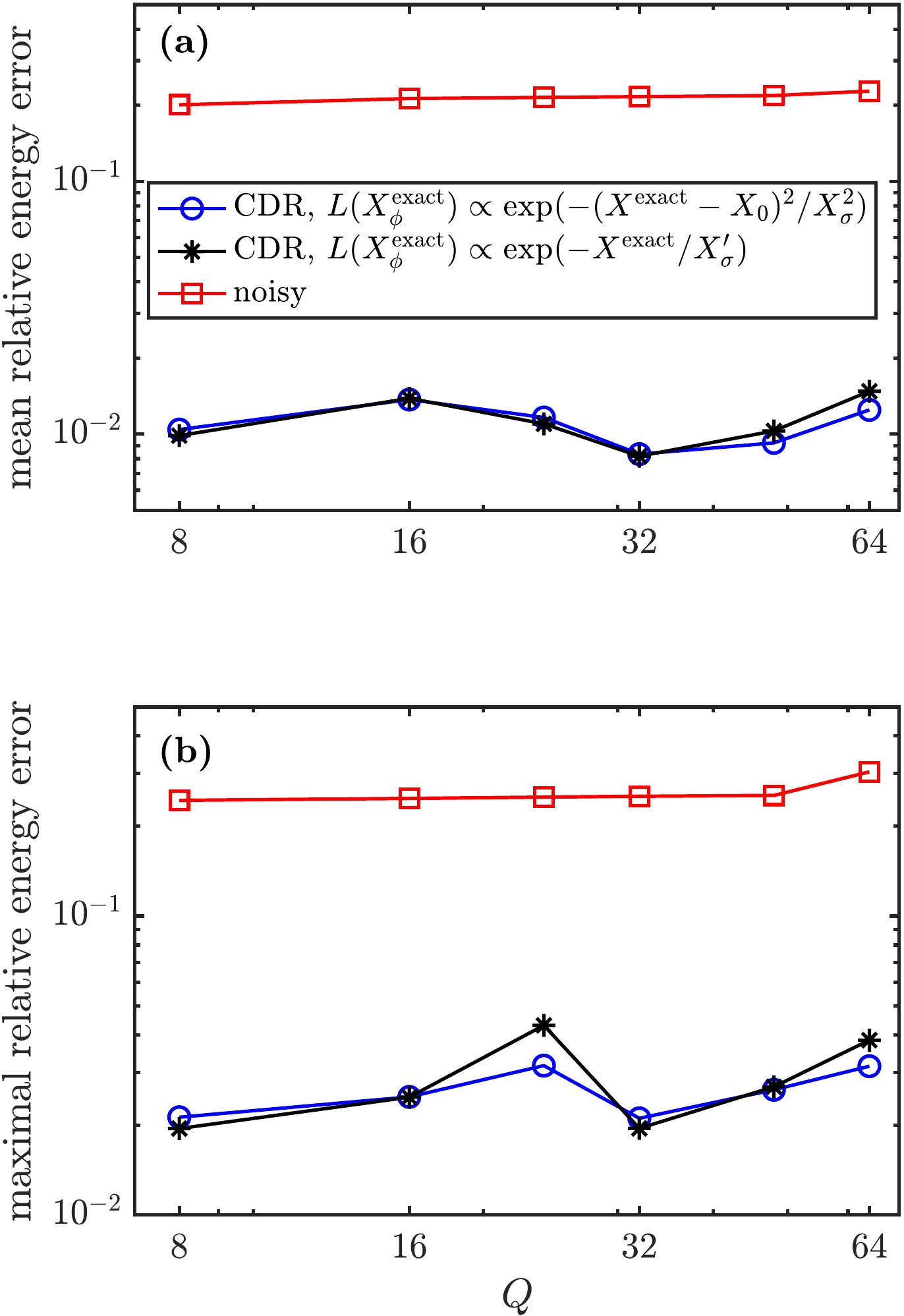}
    \caption{ Comparing the quality of the correction for different MCMC likelihoods $L$ used to construct training sets. Here we correct the same local minima of $Q=8-64$, $p=3$ QAOA  optimization  as in Fig.~\ref{fig:res}(d). As in Fig.~\ref{fig:res}(d) we use the IBM's Ourense noise model. The blue curves are  results obtained with the likelihood  (\ref{eq:QAOA_likelihood}) which is used in the main text and are taken from Fig.~\ref{fig:res}(d).  The black curves are obtained with the likelihood  (\ref{eq:minimization_likelihood}).  Here $X=E/Q$, $X_0=-2.1$, $X_{\sigma}=0.05$ and   $X'_{\sigma}=0.02$.  The likelihood function (\ref{eq:QAOA_likelihood}) utilizes special knowledge about ground state of the model as it peaks close to the ground state energy. That's not the case for the other one as it increases monotonically with decreasing energy. We find that both approaches yield similar quality of the mitigation showing that special knowledge about the ground state of the model is not necessary to obtain a good quality error mitigation with CDR. In both cases we use $70$ training circuits and $1.16\times10^6$ shots for the mitigation.              }
    \label{fig:comp_MCMC}
\end{figure}

For  this proof of principle demonstration we chose $X_0$  close to the ground state energy. Nevertheless, for open problems we may lack knowledge about the optimization solution to base choice of $X_0$ on that.  One may set $X_0$ using lower bounds on the  energy or performing MCMC sampling  with  decreasing $X_0$ until training circuits energy distribution stops shifting towards smaller energies. 
 An arguably more elegant and simpler solution is using likelihood function that increases monotonically with decreasing energy. An example of such a likelihood is     
\begin{equation}\label{eq:minimization_likelihood}
L(X^{\exact}_{\phi}) \propto e^{-X^{\exact}/X'_\sigma}.
\end{equation}
In Fig.~\ref{fig:comp_MCMC} we test this likelihood for correcting energy for local minima of $Q=8-64$, $p=3$ QAOA optimization.  These circuits were corrected using the likelihood (\ref{eq:QAOA_likelihood}) in Fig.~\ref{fig:res}(d). As in Fig.~\ref{fig:res}(d),  we use the noise model of IBM's Ourense quantum computer. Furthermore, we choose the same number of training circuits and total shots spent performing error mitigation as for simulations from Fig.~\ref{fig:res}(d).   With $X'_{\sigma}=0.02$ we find that the quality of the correction  is similar to the one obtained by earlier approach. This result demonstrates that knowledge of the solution's properties is not necessary to construct a good training set.  The value of $X'_{\sigma}=0.02$ is chosen to ensure that the exact energies of the training circuits are close to the lowest energies obtained with the training circuits constructed by randomly projecting the non-Clifford gates according to (\ref{projloc}).  

For the phase estimation example, we used a similar expression to what we used for QAOA:
\begin{equation}\label{eq:pe_likelihood}
L(X^{\noisy}_{\phi}) \propto e^{-(X^{\noisy}_{\phi}-X^{\noisy}_{\psi})^2/X_\sigma^2},
\end{equation}
with $X_\sigma=0.1$. This version is appropriate when correcting observables that we do not have special information about (e.g., that $X$ is not supposed to be minimized).

To give insight into resource cost  of the CDR method  we analyze in detail construction of the training sets for deep QAOA circuits from Fig.~\ref{fig:QAOA_deep} and large, shallow circuits from  Fig.~\ref{fig:res}(d). The MCMC procedure described above generates a chain of near-Clifford circuits indexed by an index $i$ for each mitigated circuit. We build the training set  choosing the chain  elements with  $i=i_0+k\xi_{\textrm{MCMC}}, \, k=0,1,\dots,69$, where $\xi_{\textrm{MCMC}}$ measures auto-correlation in the chain and $i_0$ determines how many initial elements of the chain are rejected. We define $\xi_{\textrm{MCMC}}$  as 
\begin{equation}
\frac{
\sum_{i=i_0}^{L-\xi_{\textrm{MCMC}}} (\vec{\theta}_i - \bar{\vec{\theta}}) \cdot (\vec{\theta}_{i+\xi_{\textrm{MCMC}}} - \bar{\vec{\theta}})}
{
\sum_{i=i_0}^{L-\xi_{\textrm{MCMC}}} (\vec{\theta}_i - \bar{\vec{\theta}}) \cdot (\vec{\theta}_{i} - \bar{\vec{\theta}})} = \frac{1}{10}
\label{xi_MCMC} \ ,
\end{equation}
where $\vec{\theta}$  is a vector of $R_Z$ rotations' angles  in the circuit, $\bar{\vec{\theta}} = \frac{1}{L-i_0}\sum_{i=i_0}^L \vec{\theta}_i$ and $L$ is the chain's length.
For fixed $N$, required classical resources are determined in our case  by $\xi_{\textrm{MCMC}}$, $i_0$ and number  of  Clifford gates in the circuit.

\begin{figure}[h!]
    \centering
    \includegraphics[width=1.\columnwidth]{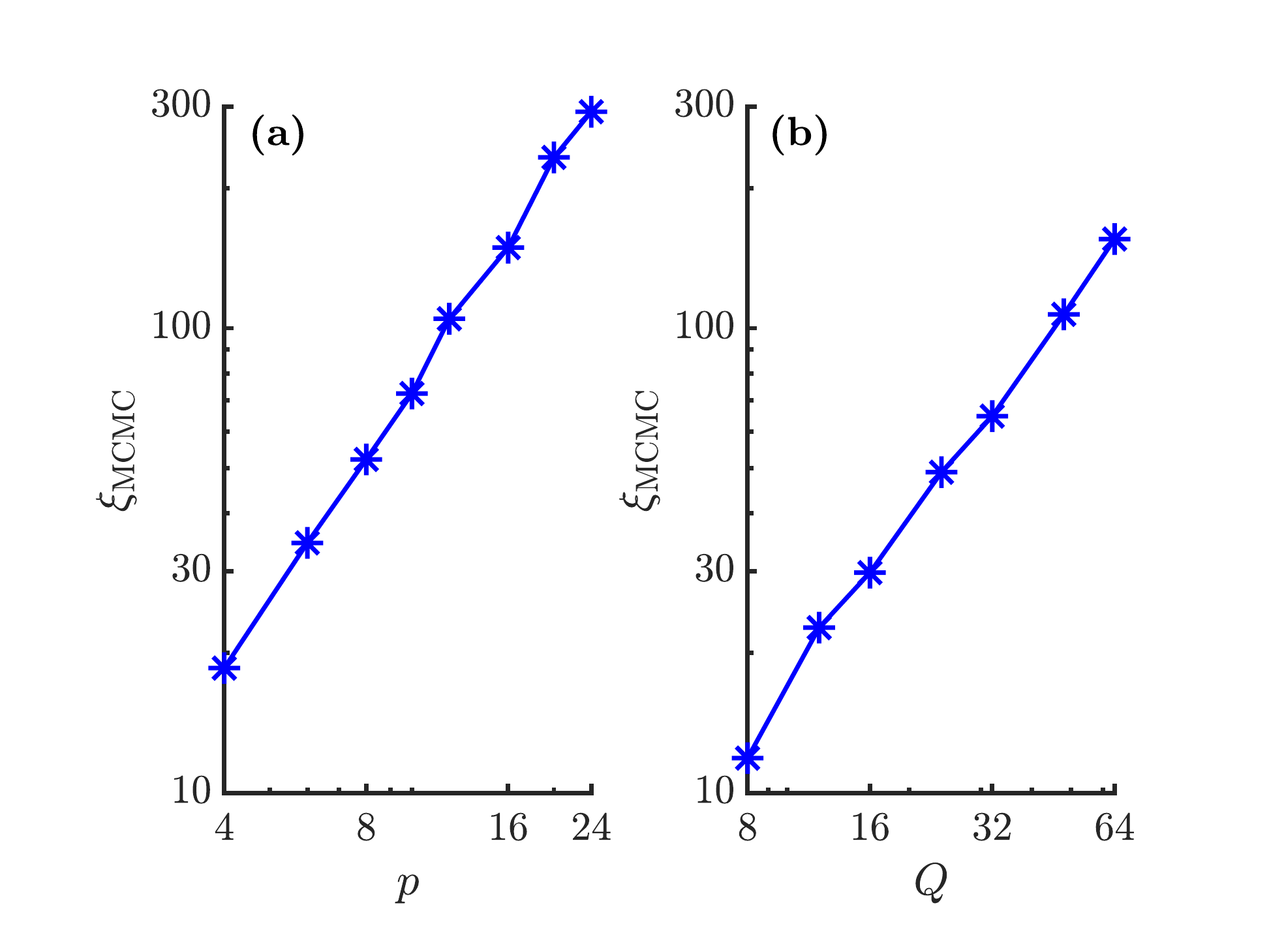}
    \caption{ 
     Classical computational cost of the training set construction.  In (a) the mean value of a MCMC auto-correlation length $\xi_{\textrm{MCMC}}$ (\ref{xi_MCMC}) for deep QAOA circuits from Fig.~\ref{fig:QAOA_deep} is plotted versus $p$ using log-log scale. In (b) the same for large shallow circuits from  Fig.~\ref{fig:res}(d). $\xi_{\textrm{MCMC}}$ is one of the factors determining cost of the classical MCMC simulations, see  text for a detailed explanation.   It appears to scale polynomially with increasing number of QAOA rounds $p$ and number of qubits  $Q$.  
    }
    \label{fig:xi_MCMC}
\end{figure}

The classical resources scale polynomially  with increasing number of the Clifford gates. Scaling of  $\xi_{\textrm{MCMC}}$ and $i_0$ needs to be analyzed empirically.   We observe that mean value of $\xi_{\textrm{MCMC}}$ obtained by averaging over the minima scales polynomially with increasing $p$ and $Q$, see Fig.~\ref{fig:xi_MCMC}.
The behavior of $i_0$ is somewhat more complicated as it depends on a starting point of the MCMC chain. If the initial point with the energy close to the chain's stationary energy distribution, one can choose $i_0=1$ without decreasing quality of the energy correction. Otherwise it is necessary to reject some initial elements of the chain. For the shallow circuits and the deep one with $p\le16$  we were able to find  good initial points starting from a set of $200$ near-Clifford circuits obtained by our chain initialization algorithm. That is no longer possible for $p=20,24$ for which we choose $i_0=70\xi_\textrm{MCMC}+1$ and $i_0=150\xi_\textrm{MCMC}+1$, respectively.   We remark that to reduce  classical computation time for these $p$ values one can choose the initial points using MCMC chains obtained for smaller $p$. To sum up we observe that for fixed $N$  total cost of classical simulations for our implementation scales polynomially with $Q$ and $p$  while showing  deviations for the largest  $p$. Furthermore, we note that the MCMC sampling for $N=28$ and the largest circuits ($Q=8, p=24$ and $Q=64, p=3$) takes less than  $2$ days  and $12$ hours, respectively,   while using a single core of  Intel Xeon E5-2695 (v4,  2.1 GHz) processor per a circuit of interest.

\section{Error bars}
\label{Ap:bars}

In the main text, we display error bars on the corrected observables obtained by the CDR method. These error bars are meant to convey one's confidence in the predicted noise-free observable. Conceptually speaking, there are two main sources of error to consider: (1) Imperfect training on the training data such that the cost function $C$ does not go to zero, and (2) Inability of the training data to capture the noise processes that affect the target state $\ket{\psi}$. The latter source of error is difficult to quantify in practice, although it can be systematically removed by increasing the number of non-Clifford gates $N$ as discussed in the main text. 

Therefore we focus our error bars on the former source of error, i.e., imperfect training. For the most part, we find (see Fig.~4) that the error bars associated with imperfect training are sufficient to encompass the discrepancy between the predicted and exact observable values. However, for our largest implementation, $Q=64$ in Fig.~\ref{fig:res2}(d), one can see that our error bars sometimes underestimate the true error. This suggests that our error bars are useful as lower bounds on the true error.

As mentioned in the main text, we calculate our error bars using the value of the cost function $C$ obtained after training. Specifically, the magnitude of the error bar is given by three standard deviations, where the standard deviation is given by $\sqrt{C/(L-1)}$, where $L$ is the number of states in the training set $\SC_{\psi}$.

\section{QAOA circuit structure}
\label{Ap:QAOA}

\begin{figure}[t]
\includegraphics[width=\columnwidth,clip=true]{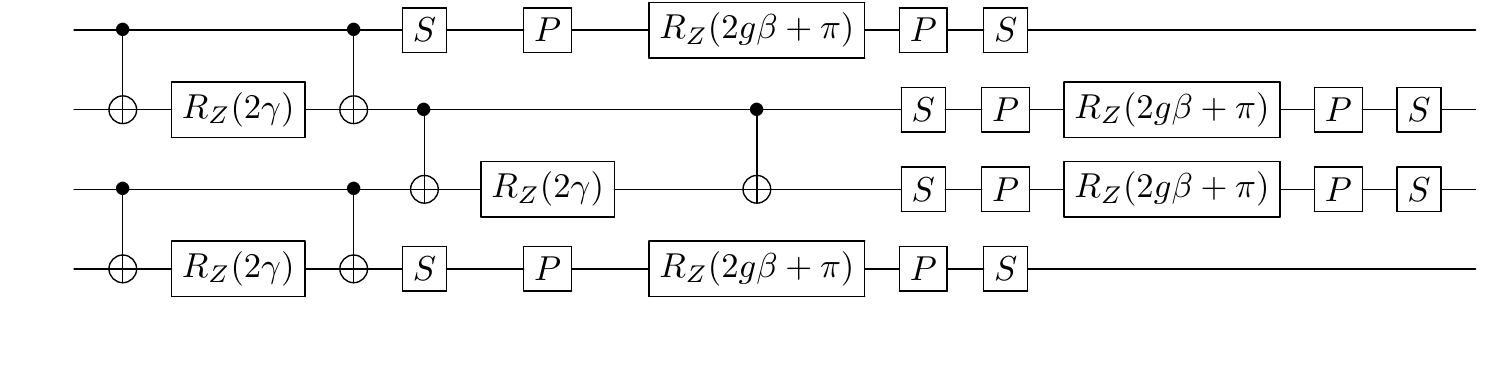}
\vspace{-0cm}
\caption{A circuit implementing a layer of the QAOA ansatz for a system with $Q=4$ qubits. $\beta, \gamma$ are QAOA parameters. Here $P=R_X(\pi/2)=e^{-i\sigma_{X}\pi/4 }$ and is a Clifford gate. The only non-Clifford gates in this circuit are the $\sigma_Z$ rotations: $R_Z(\alpha) = e^{-i\sigma_Z\alpha/2}$. We note that we have made use of the decomposition of $e^{i\gamma \sigma_{Z}^{j} \sigma_{Z}^{j'}}$ from Ref.~\cite{LANLqalgs}.}
\label{fig:qaoa}
\end{figure}

Figure~\ref{fig:qaoa} shows the structure of our QAOA circuit for the Ising spin chain considered in the main text. We note that all gates except for the the $(2Q-1)p$ $R_Z$ gates are Clifford gates. The circuits contain $(2Q-2)p$ CNOTs.   The only changes made to this circuit for generating the training data $\SC_{\psi}$ are therefore only replacements of these $\sigma_Z$ rotations by $S^n$, as discussed in Appendix~\ref{Ap:MCMC}.

\begin{figure*}[t]
\includegraphics[width=0.85\textwidth,clip=true]{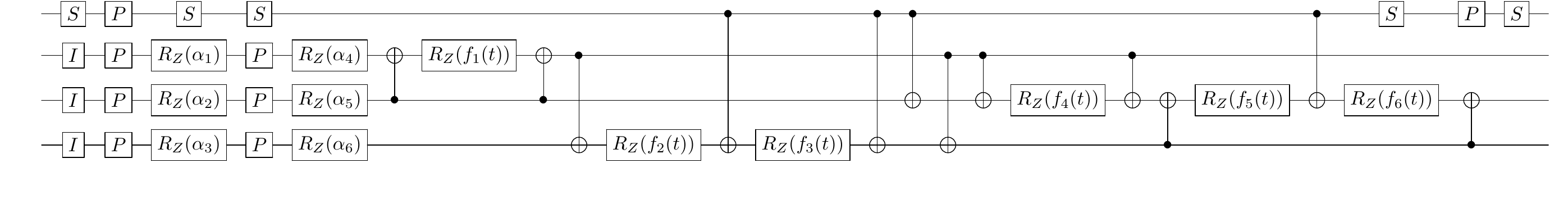}
\caption{
Quantum circuit used to estimate $\Re(\mte{\chi}{e^{-i\tilde{H}t}})$ for $\tilde{H} $, and a random product state $\ket{\chi}$ given by  random angles $\alpha_1,\alpha_2,\alpha_3,\alpha_4,\alpha_5,\alpha_6$.   Parameters $f_1(t)$, $f_2(t)$, $f_3(t)$, $f_4(t)$, $f_5(t)$, $f_6(t)$ were obtained by compiling the circuit  with an algorithm from Ref.~\cite{cincio2018learning}.  The compilation was accurate up to numerical precision. We assume here all to all device connectivity.}   
\label{fig:circphase}
\end{figure*}

\section{Quantum Phase Estimation circuit structure}\label{Ap:QPE}

Figure~\ref{fig:circphase} shows the quantum phase estimation circuit from Ref.~\cite{somma2019quantum} for $\tilde H= \frac{1}{6} (\sigma^{1}_Z \sigma^{2}_Z +\sigma^{1}_Z \sigma^{3}_Z + \sigma^{2}_Z \sigma^{3}_Z)$  compiled  by an algorithm from Ref.~\cite{cincio2018learning}  and applied to a randomly chosen input product state.   As in the QAOA ansatz, we note that all gates except for the the $\sigma_Z$ rotations are Clifford gates. Again, our classically simulable training data set $\SC_{\psi}$ is generated by replacing some of these $\sigma_Z$ rotations with $S^n$ as discussed in Appendix~\ref{Ap:MCMC}.

\section{Zero Noise Extrapolation}
\label{Ap:ZNE}

\begin{figure*}
    \centering
    \includegraphics[width=\textwidth]{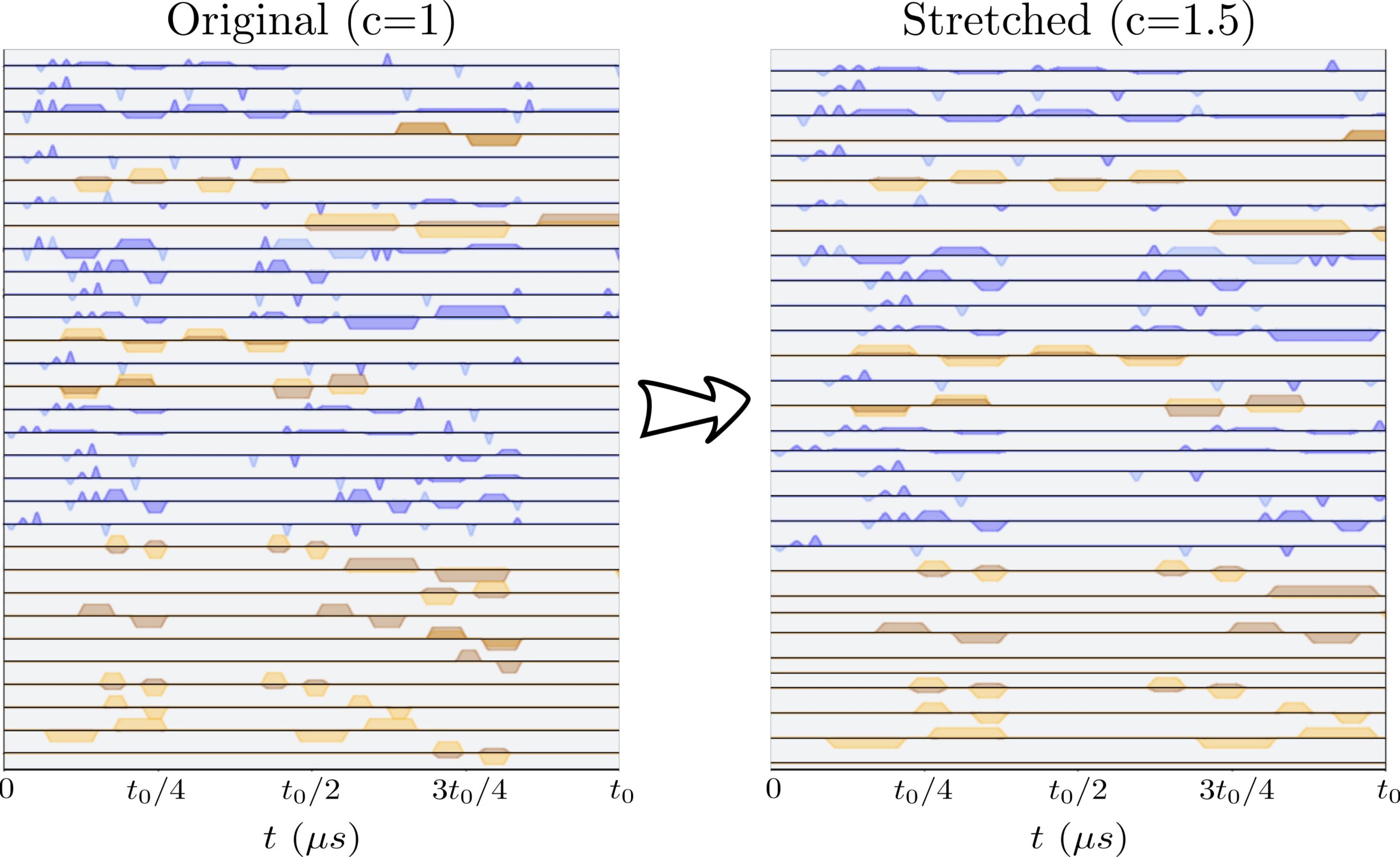}
    \caption{Stretching the QAOA circuit pulse sequence. The first $t_0 = 2.22 \,\mu s$ of the pulse sequences for a  QAOA circuit with $Q=16$ qubits and $p=2$ QAOA rounds and stretch factors  $c=1$ and $c=1.5$ are shown. In the stretched case the pulse envelopes are lower amplitude but longer in duration. For convenience, the amplitude of the pulse envelopes shown have been normalized between the channels (shown as horizontal lines), but the normalizations are the same between the $c=1$ and $c=1.5$ cases.}
    \label{fig:pulses}
\end{figure*}

\begin{figure*}
    \centering
    \includegraphics[width=0.6\textwidth]{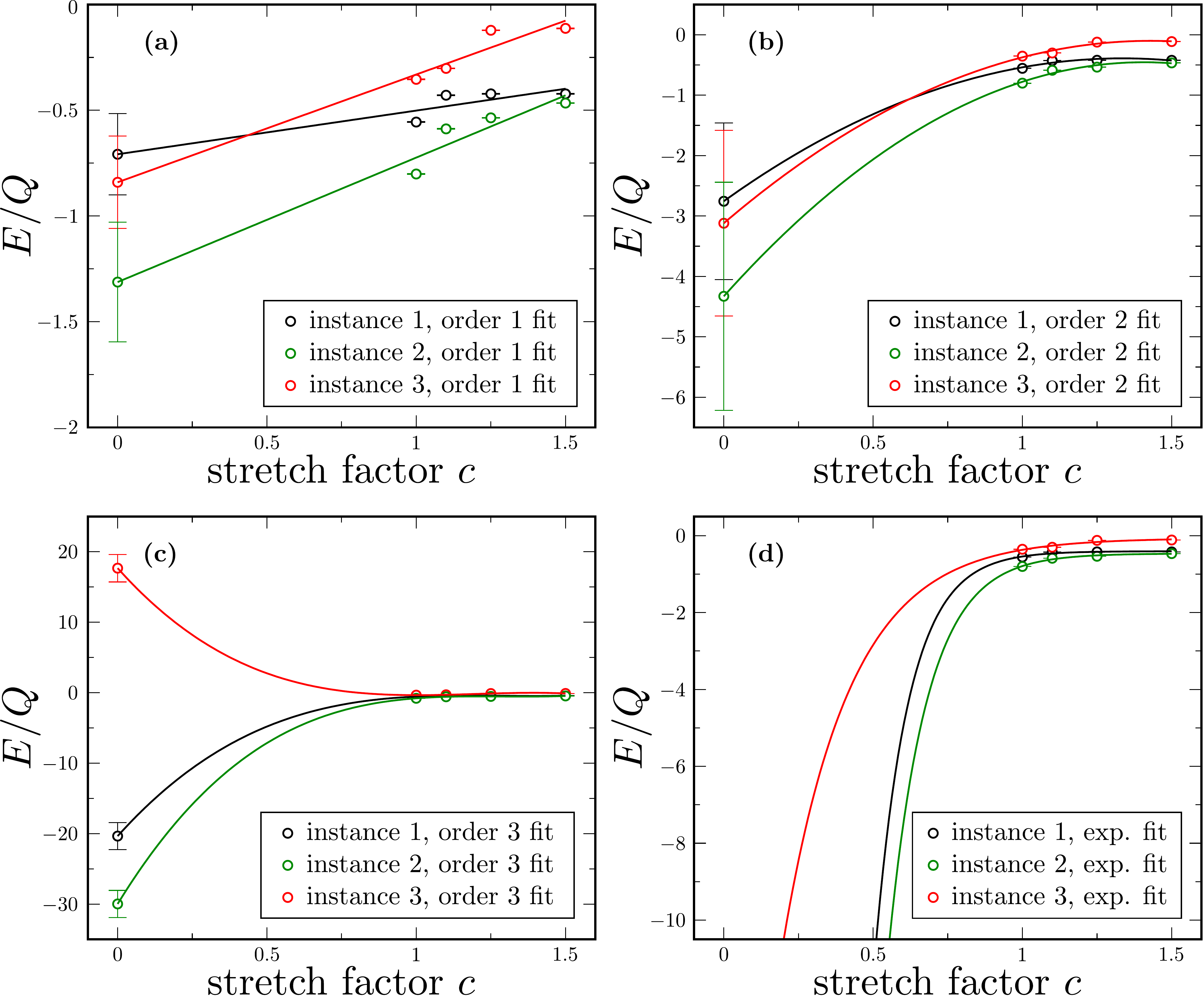}
    \caption{Extrapolating the stretched circuit data. Note that the error bars displayed here represent the uncertainties of the fit for $c=0$ and the uncertainty from finite statistics for $c\in \{1,1.1,1.25,1.5 \}$. (The finite statistics error bars are much smaller than the markers.)  Note also that the third order fit is equivalent to using equation (3) in~\cite{temme2017error}. For reference, the true values of $E/Q$ for instances 1, 2, and 3 are roughly $-2.1153$, $-2.1045$, and $-2.0990$, respectively. Corresponding  CDR mitigated vales of  $E/Q$ are $-2.071(54)$, $-2.038(26)$, $-2.048(39)$, respectively, with  the parentheses being CDR error bars.    These instances correspond to the first three instances shown in Fig.~\ref{fig:IBMQresults}.}
    \label{fig:ZNE_plots}
\end{figure*}

For comparison with our CDR method, we have performed zero noise extrapolation (ZNE) for the QAOA implementations discussed in the main text on IBM's 20-qubit Almaden quantum processor.    Specifically, we used Qiskit's Pulse package~\cite{Qiskit} to systematically stretch the microwave pulse sequences used to physically implement our $Q=16$, $p=2$ optimized QAOA circuits as done in Ref.~\cite{kandala2019error}. 

\subsection{Background}

Following Ref.~\cite{temme2017error}, the evolution of the quantum computer can be modeled in terms of the drive Hamiltonian described by the quantum circuit, $K(t)$ and a Lindblad operator $\LC\left(\rho\right)$ representing the physical noise channels:
\begin{equation}\label{eq:Lindblad}
    \frac{\partial}{\partial t}\rho(t)= -i \left[K(t),\rho(t) \right] + \lambda \LC\left(\rho(t)\right).
\end{equation}
Here $\lambda$ is a (hopefully) small parameter that represents the strength of the action of the noise channels, so the limit $\lambda \to 0$ would represent noiseless quantum computation. Attempting to approximate this limit is the heart of ZNE. While an experimenter cannot in general directly adjust $\lambda$, under the assumption that the form of $\LC$ is invariant under time re-scaling and independent of $K(t)$, one can in effect increase $\lambda$.  

Increasing $\lambda$ is accomplished by increasing (stretching) the time for the circuit evolution ($T$) by a factor of $c$ while decreasing the magnitude of the drive Hamiltonian:
\begin{equation} \label{eq:rescaling}
\begin{split}
    T&\to T'=cT,\\
    K(t)&\to K'(t)=c^{-1}K(c^{-1}t).
\end{split}
\end{equation}
To see how this works, let us integrate \eqref{eq:Lindblad} with respect to time from $t=0$ to $t=T'$, using our modified drive Hamiltonian from \eqref{eq:rescaling}. Calling the state evolved this way $\rho'(t)=\rho(c^{-1}t)$, we have:
\begin{align} \label{eq:rescaled_rho}
    &\rho'(T')=\rho(0) -i \int_0^{cT}\left[K'(t),\rho'(t) \right]dt + \lambda \int _0^{cT}\LC\left(\rho'(t)\right)dt\nonumber\\
    &=\rho(0) -i \int_0^{T'}\left[K(t),\rho(t') \right]dt' + c\lambda \int _0^{T'}\LC\left(\rho(t')\right)dt'.
\end{align}
We therefore have that, under these assumptions, the final state driven over a longer time with the stretched drive Hamiltonian is equivalent to one evolved with the original drive Hamiltonian with $\lambda\to c\lambda$. For a more detailed derivation of this formalism, see Ref.~\cite{temme2017error}.

\subsection{Implementation}

For our implementation, we follow Ref.~\cite{kandala2019error} in choosing the stretch factors $c\in \{1,1.1,1.25,1.5 \}$. We then stretched the pulse sequences generated from the QAOA circuit as shown in Fig.~\ref{fig:pulses}. In addition to running the circuit at different stretch factors, we also made use of Qiskit's built-in measurement error mitigation functions~\cite{Qiskit} as ZNE does not directly handle read-out error. Finally, we used $212992$ shots to measure each operator for each value of $c$.

\subsection{Results}

The data points we measured as well as our extrapolated energies are shown in Fig.~\ref{fig:ZNE_plots} for the three instances of low energy QAOA studied in the main text. For the sake of completeness, we show the extrapolation with a linear fit, a quadratic fit, and finally the cubic polynomial that is the standard ZNE approach for four values of $c$~\cite{temme2017error,kandala2019error}. In addition, we also show an exponential fit for ZNE~\cite{endo2018practical}. As shown in Fig.~\ref{fig:ZNE_plots}, it appears that for our particular use case the ZNE method did not provide an accurate correction for the energy expectation values.

\section{Correction with a constant ansatz}

\label{Ap:constant}

Training set construction methods tested in the paper are reliant on building a set of training circuits similar in some respect to the circuit of interest. In the case of the QAOA error mitigation from Section~\ref{sec:QAOA} the training circuits are chosen to have similar energy as the circuit of interest. In the case of Quantum Phase Estimation presented in Section~\ref{sec:QPE}  they are chosen to have  similar noisy expectation value of the mitigated observable. One may wonder if good performance of the method is limited to the cases for which  the exact expectation value of the mitigated observable is very similar for both the training circuits and the circuit of interest.

\begin{figure}[t!]
\includegraphics[width=0.9\columnwidth,clip=true]{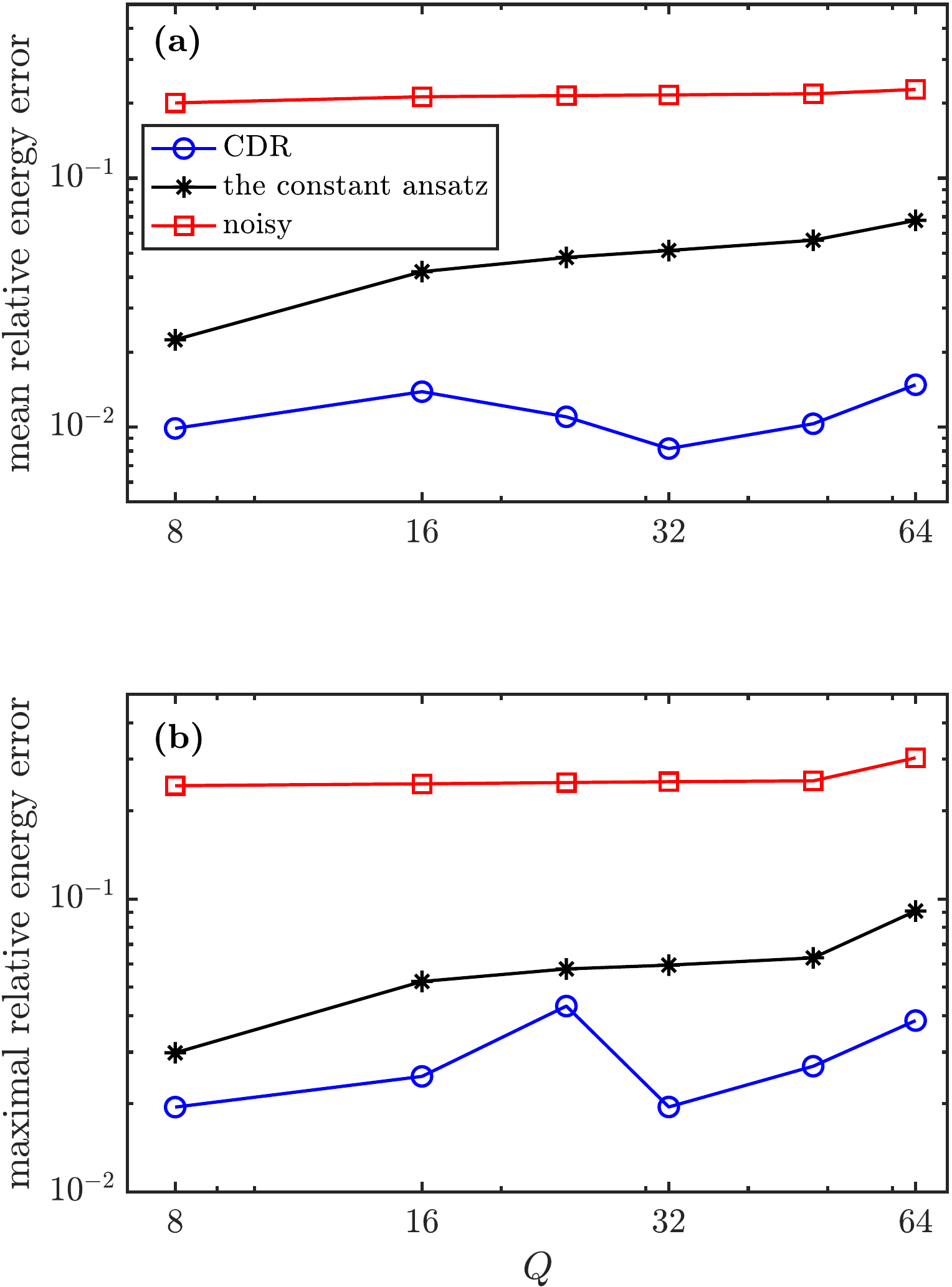}
\caption{
 Quality comparison of the CDR correction  and the  constant ansatz (\ref{eq:constant}) correction. Here we correct the same local minima of $Q=8-64$, $p=3$ QAOA  optimization  as in Figs.~\ref{fig:res}(d),~\ref{fig:comp_MCMC} of the manuscript. We  use the IBM Ourense noise model.  The blue curves show the results obtained with the CDR and are taken from Fig.~\ref{fig:comp_MCMC}.  The black curves depict the results obtained with the constant ansatz using  the same training sets as for CDR. We see that the CDR  outperforms constant correction performed with (\ref{eq:constant}). The difference is especially pronounced for  circuits with the largest qubit count. }   
\label{fig:constant}
\end{figure}

In the case of simple  global depolarizing noise treated analytically in Appendix~\ref{Ap:LinearAnsatz}, we see that  CDR is capable of learning the  perfect correction from any two  training circuits with different values of the observable of interest. Therefore, at least in some cases  the method has potential to learn correction even if the training circuits  have expectation values  of the observable of interest very different from the circuit of interest. In this appendix we give insight into this  problem  for the more realistic case of the IBM Ourense noise model. To that end we  perform the correction with a constant ansatz 
\begin{equation}
f(X^{\noisy}_{\psi}, \vec{a} )= a_1
\label{eq:constant}
\end{equation}
and compare it with  CDR correction. 
As a test case  we  correct  the $p=3$ QAOA circuits for which CDR correction was performed in  Figs.~\ref{fig:res}(c) and~\ref{fig:comp_MCMC}.   We  use the training sets used above  in  Fig.~\ref{fig:comp_MCMC}  which are constructed with MCMC sampling method~(\ref{eq:QAOA_likelihood}). We present the results in Fig.~\ref{fig:constant} showing that CDR  outperforms the correction with the constant ansatz. This shows that good quality of the correction can be obtained for the realistic noise even when construction of the training set with $X_{\phi_i}^{\exact}$ concentrated around  $X_{\psi}^{\exact}$ is impossible. 
Finally, to provide additional insight into properties of the used training sets we gather average absolute values of  the fitted CDR ansatz coefficients in Tab.~\ref{tab:coeff}.

\begin{table}[t!]
\begin{tabular}{|c|c |c |}
\hline
 $Q$ & $a_1$  & $a_2$ \\
\hline
8 & 1.06(18) & 0.20(7) \\
16 & 0.98(21) & 0.28(9) \\
24 & 0.81(21) & 0.28(9) \\
32 & 0.79(13) &  0.27(7) \\
48 & 0.75(14) & 0.26(7) \\
64 & 0.90(16) & 0.17(6) \\
\hline
\end{tabular}
\caption{
Average absolute  values of fitted slope $a_1$ and  intercept $a_2$ of the CDR ansatz~(\ref{regression}) used to perform error mitigation in Fig.~\ref{fig:constant}. We average  over different local minima of the QAOA optimization and different  observables contributing to the energy. We also include standard deviation of the absolute values using parentheses notation. 
}   
\label{tab:coeff}
\end{table}

\clearpage


\end{document}